\documentclass[12pt]{article}
\usepackage{amsmath,centernot}
\usepackage{amsfonts}
\usepackage{amsthm}
\usepackage{graphicx}
\usepackage{enumerate}
\usepackage{natbib}
\usepackage{url} % not crucial - just used below for the URL 
\usepackage{tikz}
\usepackage{float}
\usepackage{subfigure}
\usepackage{enumitem}
\usepackage{adjustbox}

\newcommand{\indep}{\rotatebox[origin=c]{90}{$\models$}}

\newcommand{\nindep}{\centernot{\indep}}

\newtheorem{theorem}{Theorem}[section]

\newtheorem{assumption}{Assumption}
\newtheorem{example}{Example}

\begin{document}

\title{\bf Proximal mediation analysis}
\author{Oliver Dukes$^1$, Ilya Shpitser$^2$ and Eric Tchetgen Tchetgen$^1$\\
$^1$: University of Pennsylvania, Philadelphia, Pennsylvania, U.S.\\
$^2$: Johns Hopkins University, Baltimore, Maryland, U.S.}
\date{\today}

  \maketitle

\begin{abstract}
A common concern when trying to draw causal inferences from observational data is that the measured covariates are insufficiently rich to account for all sources of confounding. In practice, many of the covariates may only be \textit{proxies} of the latent confounding mechanism. Recent work has shown that in certain settings where the standard `no unmeasured confounding' assumption fails, proxy variables can be leveraged to identify causal effects. Results currently exist for the total causal effect of an intervention, but little consideration has been given to learning about the direct or indirect pathways of the effect through a mediator variable. In this work, we describe three separate proximal identification results for natural direct and indirect effects in the presence of unmeasured confounding. We then develop a semiparametric framework for inference on natural (in)direct effects, which leads us to locally efficient, multiply robust estimators.

 %In this work, we build on recent developments in \cite{tchetgen2020introduction}, who show how proxy variables can be leveraged to identify causal effects in settings where the standard `no unmeasured confounding' assumption would fail.

% In particular, we consider identification and estimation of natural direct and inter effects, in studies where one wishes to learn about the mechanism 
\end{abstract}


\section{Introduction}

%In the last two decades, there has been much development 
The last few decades has seen the emergence of a literature on causal mediation analysis  \citep{robins1992identifiability,pearl2001direct,vanderweele2009conceptual,imai2010general,tchetgen2012semiparametric}. This literature provides nonparametric definitions of direct and indirect effects in terms of contrasts of potential outcomes, as well as conditions necessary to identify and estimate these effects from data. %This line of research originated in \citet{robins1992identifiability} and \citet{pearl2001direct},  which in turn generalises earlier work in the structural equation modelling framework \citep{baron1986moderator}, to allow for possible nonlinearities and interactions in models for the outcome and/or mediator. 
Estimands that have received particular focus are \textit{natural direct and indirect effects}, which are useful for understanding the mechanism underlying the effect of a particular intervention as they combine to produce the total causal effect. 

The majority of work on identification of natural direct and indirect effects assumes that the measured covariates are sufficiently rich to account for confounding between the exposure and outcome, the mediator and outcome and the exposure and mediator. In practice, it is likely that many key confounding variables (e.g. disease severity, socio-economic status) cannot be ascertained with certainty from the measured covariates.  At best, some of the measured covariates may be confounder \textit{proxies} e.g. mis-measured versions of the underlying confounders. This insight has led to work on leveraging proxy variables to help remove confounding bias in observational studies, with focus on the total effect of intervention. Negative control are examples of such proxies \citep{lipsitch2010negative,shi2020selective}; we refer to \citet{shi2020selective} and \citet{tchetgen2020introduction} for further examples in observational studies. 

 %; initial work focused on identification in the presence of unmeasured confounding with either a negative control exposure or outcome under restrictions such as linearity \citep{flanders2011method,gagnon2012using,wang2017confounder}, rank preservation \citep{tchetgen2014control},  and monotonicity \citep{sofer2016negative}. 
If we are able to collect data on a sufficient number of proxies, confounding bias can sometimes be successfully removed in settings where standard analyses under a `no unmeasured confounding' assumption would fail. Building on results in \cite{kuroki2014measurement}, \citet{miao2018identifying} established nonparametric identification of the average treatment effect under a `double negative control design' (where both a negative control exposure and outcome are measured). 
\citet{tchetgen2020introduction} extend these results to settings with time-varying exposures and (potentially unmeasured) confounders. For estimation, they propose proximal g-computation, a generalisation of Robins' parametric g-computation algorithm \citep{robins_new_1986}. %Proximal g-computation requires estimation of a so-called `confounding bridge' function, which can be understood as the transformation of the outcome-inducing proxy that reflects the effect of the confounder on the outcome. 
Under a proximal identification strategy, \citet{cui2020semiparametric} develop semiparametric inference for the average treatment effect. %They propose locally efficient, doubly robust estimators which are consistent so long as at least one of the confounding bridge models is correctly specified. 
%They obtain the efficient influence function, which requires estimation of a second confounding bridge function. Based on these results, they propose locally efficient, doubly robust estimators which are consistent so long as at least one of the confounding bridge models is correctly specified. 

\iffalse
We consider identification and estimation of natural direct and indirect effects in the presence of unmeasured confounding. As an example, we consider the JOBS II study  \citep{vinokur1995impact}. Beyond understanding the total effect of a job training intervention, the investigators were also interested in whether the intervention improved mental health status by enhancing the confidence of participants in their job searching ability. Although the intervention was randomised, it was possible that the association between mediator (job search self-efficacy) and outcome (depression symptoms) was subject to confounding by latent factors that was only partially captured by the pre-treatment covariates (e.g. questionaire scores).
\fi
We consider identification and estimation of natural direct and indirect effects in the presence of unmeasured confounding. As an example, we consider the Job Corps study  \citep{schochet2008does}. Beyond understanding the total effect of a job training intervention, the investigators were also interested in whether the intervention reduced criminal activity due to increased employment. It was possible that the association between program participation, employment and criminal activity were subject to confounding by latent factor, such as motivation, that was only partially captured by the pre-treatment covariates. In this work, we establish sufficient conditions for nonparametric identification of mediation estimands using a pair of proxy variables, giving three separate identification strategies. These each rely on modelling and estimation of different combinations of `confounding bridge' functions  \citep{miao2018identifying}. 
%Our first identification result extends the mediation formula of \citet{pearl2001direct} to the proximal learning framework. The estimator implied by this result relies on modelling and estimation of two \textit{nested} confounding bridge functions. We then outline two alternative identification strategies that incorporate different confounding bridge functions. Given that these bridge functions are only defined implicitly as the solution to integral equations, the task of correctly specifying parametric models for these quantities may be challenging.
To reduce sensitivity to model misspecification, we obtain the efficient influence function under a semiparametric model for the observed data distribution, which leads us to estimators that are multiply robust. Our identification and estimation results allow for continuous or discrete outcomes and mediators. %In this case, they are consistent so long as at least one out of three possible sets of restrictions on the observed data hold. They are also locally efficient, in the sense that they attain the semiparametric efficiency bound under the union model when all working models are correctly specified. 
As far as we are aware, this is the first paper to use proxy variables for identification and inference for direct and indirect effects, with the exception of \citet{cheng2021causal}. However, their identification strategy is distinct from ours, as they rely on deep latent variable models. They also did not consider semiparametric inference and the issues of efficiency/robustness explored here. 
%The article is organised as follows. In Section \ref{npi}, we introduce notation and review identification of natural (in)direct effects when all confounders are measured. We then outline the assumptions required for nonparametric identification using proxy variables, before outlining three different identification strategies. In Section \ref{speff}, we obtain the efficient influence function under a nonparametric model, and then use this to construct locally efficient, multiply robust estimators. We discuss how to estimate nuisance parameters indexing the bridge functions, and also provide results for mediation analyses in randomised trials. A simulation study and data analysis follows in Sections 4 and 5 respectively, and we close the paper in Section \ref{disc} with a discussion. 

\section{Nonparametric proximal identification of the mediation functional}\label{npi}

\subsection{Preliminaries}

We consider a setting where one is interested in the effect of a binary treatment $A$ on an outcome $Y$ that is mediated via a single intermediate variable $M$. We use $U$ to refer to an unmeasured, potentially vector-valued confounding variable, which may be discrete, continuous, or a combination of both types. Let $Y(a,m)$ refer to the potential outcome that would be observed for someone if they were assigned to a given treatment at level $a$ and mediator at $m$; similarly, $M(a)$ denotes the potential outcome for the mediator if treatment had taken value $a$. Then the total average treatment effect of $A$ on $Y$ can be decomposed as 
\[E\{Y(1)-Y(0)\}=E[Y\{1,M(1)\}-Y\{1,M(0)\}]+E[Y\{1,M(0)\}-Y\{0,M(0)\}]\]
The first term $E[Y\{1,M(1)\}-Y\{1,M(0)\}]$ on the right hand side of the equality is an example of a natural indirect effect, and captures the expected mean difference in $Y$ if all individuals were assigned treatment $A=1$, but the mediator was changed to the level it would take with $A=0$. The second term  $E[Y\{1,M(0)\}-Y\{0,M(0)\}]$ is a natural direct effect, and captures the effect of setting $A=1$ versus $A=0$ if everyone's mediator were at the level it would take with $A=0$. Note that $E[Y\{1,M(1)\}]=E\{Y(1)\}$ and $E[Y\{0,M(0)\}]=E\{Y(0)\}$; results on nonparametric identification and inference for these quantities in a proximal learning setting already exist in \citet{tchetgen2020introduction} and \citet{cui2020semiparametric}. We will therefore focus on the mediation functional $\psi=E[Y\{1,M(0)\}]$ in the remainder of the article. 

\begin{figure}[H]
    \centering
\begin{minipage}{.5\textwidth}
	\begin{tikzpicture}[scale=0.75]
		\node[] (r) at (0, 5)   {$(a)$};
	\node[] (a) at (0, 2)   {$A$};
	\node[] (m) at (3, 4)   {$M$};
	\node[] (y) at (6, 2)   {$Y$};
	\node[] (l) at (3,0) 	 {$L$};
	%\node[] (w) at (1,5) 	 {$W$};
	%\node[] (z) at (-1,3) 	 {$Z$};

	\path[->] (a) edge node {} (y);
	\path[->] (a) edge node {} (m);
	\path[->] (m) edge node {} (y);
	%\path[->] (u) edge node {} (z);
	%\path[->] (z) edge node {} (a);
	%\path[->] (u) edge node {} (w);
	\path[->] (l) edge node {}  (m);
	\path[->] (l) edge node {} (a);
	\path[->] (l) edge node {} (y);

	\end{tikzpicture}
  %\label{fig:test1}
\end{minipage}%
\begin{minipage}{.5\textwidth}
	\begin{tikzpicture}[scale=0.75]
		\node[] (r) at (0, 5)   {$(b)$};
	\node[] (a) at (0, 2)   {$A$};
	\node[] (m) at (3, 4)   {$M$};
	\node[] (y) at (6, 2)   {$Y$};
	\node[]  (x) at (3,0) 	 {$X$};
	\node[draw,circle]  (u) at (3,-2) 	 {$U$};
	\node[] (w) at (6,0) 	 {$W$};
	\node[] (z) at (0,0) 	 {$Z$};
	%\node[] (z) at (4,1.25) 	 {$Z$};
	%\node[] (v) at (0,4) 	 {$V$};

	\path[->] (a) edge node {} (y);
	\path[->] (a) edge node {} (m);
	\path[->] (m) edge node {} (y);
	\path[->] (u) edge node {} (a);
	\path[->] (u) edge node {} (z);
	\path[->] (x) edge node {} (a);
	\path[->] (x) edge node {} (m);
	\path[->] (x) edge node {} (y);
	\path[->] (x) edge node {} (z);
	\path[->] (x) edge node {} (w);
	%\path[->] (z) edge node {} (a);
	\path[->] (u) edge node {} (y);
	\path[->] (u) edge node {} (w);
	\path[->] (u) edge node {} (x);
	%\path[->] (u) edge node {} (z);
	%\path[->] (z) edge node {} (m);
	\path[->] (u) edge [bend right=45] (m);
	%\path[->] (w) edge node {} (y);
	%\path[->] (a) edge node {} (v);
	%\path[->] (u) edge node {} (v);
	%\path[->] (v) edge node {} (m);
	\end{tikzpicture}
  %\label{fig:test2}
\end{minipage}
\caption{$(a)$ a DAG with measured confounder $L$. $(b)$ a DAG with treatment, proxies and unmeasured confounders.}
\label{standard_proxy_med}
\end{figure}
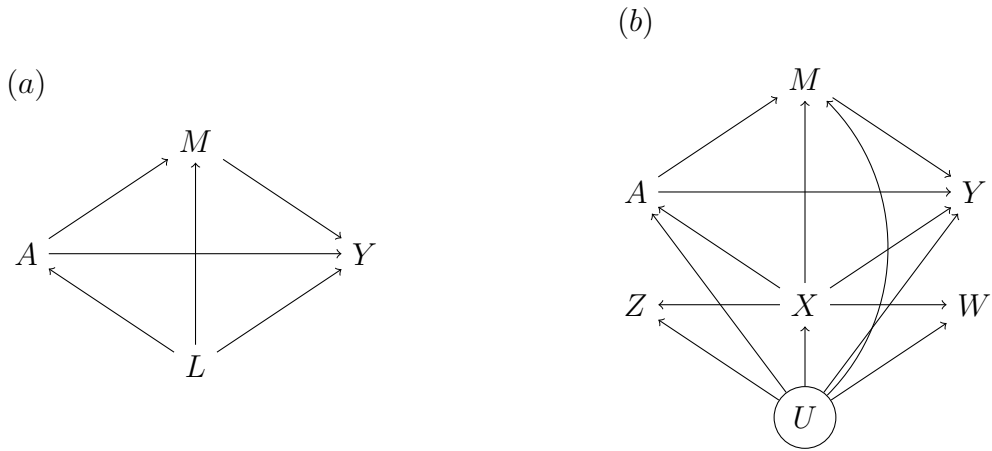

In order to identify $\psi$ when one has access to a measured, potentially vector-valued covariate $L$, and supposing $M$ takes on values in $\mathcal{S}$, then 
 one typically invokes the following conditional exchangeability assumptions: $Y(a,m)\indep A|L$ for $a=0,1$ and each $m\in \mathcal{S}$; $M(a)\indep A|L$ for $a=0,1$; and $Y(a,m)\indep M(a)|A=a,L$ for $a=0,1$ and each $m\in \mathcal{S}$. In addition, the \textit{cross-world assumption} $Y(a,m)\indep M(a')|A=a,L$ for $a,a'=0,1$ and each $m\in \mathcal{S}$ is usually invoked \citep{robins2010alternative}. It is known as such because independence between the counterfactual outcome and mediator values is required to hold across two different worlds (of potentially conflicting values of treatment). If these hold, in addition to standard positivity and consistency conditions \citep{robins_new_1986}, then $\psi$ can be identified via the \textit{mediation formula} 
\begin{align}\label{med_form}
\psi=\int \int E(Y|A=1,m,l)dF(m|A=0,l)dF(l)
\end{align}
\citep{pearl2001direct}. If one were to interpret the causal diagram in Figure \ref{standard_proxy_med}(a) as a nonparametric structural equation model with independent errors, then the above conditional independences are consistent with that diagram. 

The cross-world assumption has been the subject of much controversy, given that it can never be empirically verified or guaranteed by any study design. This has therefore led to an alternative way of conceptualising natural direct and indirect effects via `treatment-splitting' without reference to cross-world counterfactuals \citep{robins2010alternative,robins2020interventionist}.  In what follows, we will adopt the more traditional cross-world framework for mediation analysis, but expect that all of our results for identification of $\psi$ carry over to the split-treatment approach, which is left to future work.  %We will make the dismissable components assumptions throughout, and refer the interested reader to \citet{robins2010alternative}, \citet{robins2020interventionist} and \citet{stensrud2020generalized} for more information. 

\iffalse

\begin{figure}
    \centering
	\begin{tikzpicture}[scale=0.5]
	%	\node[] (r) at (0, 5)   {$(a)$};
	\node[] (a) at (0, 2)   {$A$};
	\node[] (m) at (3, 4)   {$M$};
	\node[] (y) at (6, 2)   {$Y$};
	\node[] (l) at (3,0) 	 {$L$};
	%\node[] (w) at (1,5) 	 {$W$};
	%\node[] (z) at (-1,3) 	 {$Z$};

	\path[->] (a) edge node {} (y);
	\path[->] (a) edge node {} (m);
	\path[->] (m) edge node {} (y);
	%\path[->] (u) edge node {} (z);
	%\path[->] (z) edge node {} (a);
	%\path[->] (u) edge node {} (w);
	\path[->] (l) edge node {}  (m);
	\path[->] (l) edge node {} (a);
	\path[->] (l) edge node {} (y);

	\end{tikzpicture}
\caption{a DAG with measured confounder $L$}
\label{standard_med}
\end{figure}
\fi

\subsection{The proximal mediation formula}

Figure \ref{standard_proxy_med}(b) displays a setting where the previous conditional exchangeability and the cross-world assumptions would not hold due to the presence of the unmeasured variable $U$, which is a common cause of $A$, $M$ and $Y$. We will now assume that the observed covariates $L$ can be divided into three buckets $(X,Z,W)$. %, it follows that $Y^{a,m} \nindep A|L$ where $L=(X,W,Z)$. 
Here, $Z$ and $W$ are proxy variables that are only associated with $A$, $M$ and $Y$ via an unmeasured common cause. Lastly, $X$ is a common cause of $A$, $M$ and $Y$. The failure of conditional exchangeability means that an analysis which would adjust for $X$, $Z$ and $W$ via conditioning on them in the mediation formula would return biased results due to residual confounding from $U$.
\iffalse
\begin{figure}[H]
    \centering
	\begin{tikzpicture}[scale=1]
%		\node[] (r) at (0, 4)   {$(b)$};
	\node[] (a) at (0, 2)   {$A$};
	\node[] (m) at (3.5, 4)   {$M$};
	\node[] (y) at (7, 2)   {$Y$};
	\node[]  (x) at (3.5,0) 	 {$X$};
	\node[draw,circle]  (u) at (3.5,-2) 	 {$U$};
	\node[] (w) at (7,0) 	 {$W$};
	\node[] (z) at (0,0) 	 {$Z$};
	%\node[] (z) at (4,1.25) 	 {$Z$};
	%\node[] (v) at (0,4) 	 {$V$};

	\path[->] (a) edge node {} (y);
	\path[->] (a) edge node {} (m);
	\path[->] (m) edge node {} (y);
	\path[->] (u) edge node {} (a);
	\path[->] (u) edge node {} (z);
	\path[->] (x) edge node {} (a);
	\path[->] (x) edge node {} (m);
	\path[->] (x) edge node {} (y);
	\path[->] (x) edge node {} (z);
	\path[->] (x) edge node {} (w);
	\path[->] (z) edge node {} (a);
	\path[->] (u) edge node {} (y);
	\path[->] (u) edge node {} (w);
	\path[->] (u) edge node {} (x);
	%\path[->] (u) edge node {} (z);
	%\path[->] (z) edge node {} (m);
	\path[->] (u) edge [bend right=45] (m);
	\path[->] (w) edge node {} (y);
	%\path[->] (a) edge node {} (v);
	%\path[->] (u) edge node {} (v);
	%\path[->] (v) edge node {} (m);
	\end{tikzpicture}
\caption{a DAG with treatment and outcome-inducing proxies.}
\label{proxy_med}
\end{figure}
\fi
In what follows, we will therefore adopt a different approach for identification of $\psi$, based on leveraging the proxy variables $Z$ and $W$ to learn about $U$. Before giving the first identification result, we discuss the assumptions involved.

\begin{assumption}\label{consist}
(Consistency):
\begin{enumerate}
    \item  $M(a)=M$ almost surely for those with $A=a$.
    \item $Y(a,m)=Y$ almost surely for those with $A=a$ and $M=m$.
\end{enumerate}
\end{assumption}

\begin{assumption}\label{pos}
(Positivity):\\
\begin{enumerate}
    \item $f_{M|A,U,X}(m|A,U,X)>0$ almost surely for all $m \in \mathcal{S}$. 
    \item $Pr(A=a|U,X)>0$  almost surely for $a=0,1$. 
\end{enumerate}
\end{assumption}
\begin{assumption}\label{lce}
(Latent conditional exchangeability):
\begin{enumerate}
    \item $Y(a,m)\indep A|U,X$ for $a=0,1$ and each $m\in \mathcal{S}$.
    \item $Y(a,m)\indep M(a)|A=a,U,X$ for $a=0,1$ and each $m\in \mathcal{S}$.
    \item $M(a)\indep A|U,X$ for $a=0,1$ and each $m\in \mathcal{S}$. 
\end{enumerate}
\end{assumption}
\begin{assumption}\label{cwa}
(Latent cross-world assumption):\\ $Y(a,m)\indep M(a^*)|U,X$ for $a,a^*=0,1$ and each $m\in \mathcal{S}$.
\end{assumption}
These first assumptions are similar to those made in standard mediation analyses, except that they also allow for the existence of an unmeasured variable $U$. 
%The final assumption is an extension of the assumption of \cite{robins2010alternative} for settings with unmeasured confounding and can be read off the extended causal diagram Figure 2(b). It cannot be verified in a study where only data on $A$ is collected, but could be checked e.g. in a randomised trial in which $A_Y$ and $A_M$ are randomly assigned.
The following assumptions will directly enable us to leverage information from the proxy variables.
\iffalse
\begin{assumption}\label{p_proxy}
(Paired proxies): 
\begin{align*}
&W\indep A|M,U,X,\\
&W\indep A|U,X,\\
&Y\indep Z|A,M,U,X,\\
&M\indep Z|A,U,X,\\
&W\indep Z|A,U,X,\\
&W\indep Z|A,M,U,X
%Y(a,z)=Y(a)
%Y(a,m,
\end{align*} %COME BACK TO THIS
\end{assumption}
\fi
\begin{assumption}\label{p_proxy}
(Exposure- and outcome-inducing proxies): 
\begin{enumerate}
    \item $Z\indep Y(a,m)|A,M(a),U,X$ for $a=0,1$ and each $m\in \mathcal{S}$.
    \item $Z\indep M(a)|A,U,X$ for $a=0,1$.
    \item $W\indep M(a)|U,X$ for $a=0,1$.
    \item $W\indep (A,Z)|M(a),U,X$ for $a=0,1$.
\end{enumerate}
\end{assumption}

These assumptions essentially require that $A$ and $M$ have no direct causal effect on $W$, and that $Z$ has no causal effect on $M$ or $Y$. Also, $Z$ and $W$ are only associated via the unmeasured common cause $U$. %Using the graphoid axioms, one can show that Assumptions \ref{consist}, \ref{lce}, \ref{p_proxy} and \ref{p_proxy} imply that 
%\begin{align*}
%&Y\indep Z|A,M,U,X%\label{cond_ind1}\\ 
%&M\indep Z|A,U,X %\label{cond_ind2}\\
%&W\indep M|Z,A,U,X \label{cond_ind3}
%\end{align*} (see the Appendix). 
These assumptions formally encode what it means for $Z$ and $W$ to be proxies, in the sense that they become uninformative about confounding conditional on $U$. They are compatible with the causal diagram in Figure \ref{standard_proxy_med}b. However, there are many other diagrams that may be compatible with Assumptions \ref{p_proxy}.3 and \ref{p_proxy}.4, some of which are given in Figure \ref{examples} in the Supplementary Material. For example, $Z$ can be a direct cause of $A$, and $W$ can cause $Y$. In general, these assumptions are not testable given that they involve the unmeasured $U$.

Identification of the mediation functional requires existence of multiple `confounding bridge' functions \citep{miao2018confounding}.%; for example, this can amount to assuming there exists a transformation of $W$, whereby the effect of the unmeasured $U$ on $Y$ is equal to its effect on the transformed proxy. 
We also need to ensure the confounding bridge functions can be identified from the data, given that we have access to $Z$. Formalising both of these types of condition in general settings is subtle, since the confounding bridge function will be defined as the solution to an integral equation. We formalise them using the completeness conditions below: %which captures the link between the $U-W$ and $U-Y$ associations. 
%To identify the mediation functional, we also need key proxy relevance conditions along the lines of conditions (ii) and (iii) in \cite{miao2018identifying}:
\begin{assumption}\label{complete}(Completeness):
\begin{enumerate}
\item For any square-integrable function $g(u)$, if $E\{g(U)|Z=z,A=1,M=m,X=x\}=0$ for any $z$, $m$ and $x$ almost surely, then $g(U)=0$ almost surely.
\item For any square-integrable function $g(u)$, if $E\{g(U)|Z=z,A=0,X=x\}=0$ for any $z$ and $x$ almost surely, then $g(U)=0$ almost surely.
%\item For any function $g(z)$,  $E\{g(Z)|W=w,A=1,M=m,X=x\}=0$ for any $w$, $m$ and $x$ almost surely, then $g(Z)=0$ almost surely.
%\item For any function $g(z)$,  $E\{g(Z)|W=w,A=0,X=x\}=0$ for any $w$ and $x$ almost surely, then $g(Z)=0$ almost surely. 
\end{enumerate}
\end{assumption}
%Completeness conditions are able to accommodate both continuous and categorical confounders; 
Completeness is a technical condition that arises in conjunction with sufficiency in the theory of statistical inference. In causal inference, it has been invoked for identification in the context of  nonparametric regression with an instrumental variable \citep{newey2003instrumental}, where it is used as an analogue of the rank and order conditions that arise in the classical instrumental variable setup. This assumption essentially means that any variation in $U$ is associated with a form of variation in $Z$ given $A=1$, $M$  and $X$ 
  and given $A=0$ and $X$.  Further discussion of completeness in this context is left to the Supplementary Material.

We are now in a position to give our first identification result.
\begin{theorem}\label{id_or}
Suppose that there exist the following confounding bridge functions $h_1(w,M,X)$ and $h_0(w,X)$
that satisfy
\begin{align}
E(Y|Z,A=1,M,X)&=\int h_1(w,M,X)dF(w|Z,A=1,M,X)\label{fred1}\\
E\{h_1(W,M,X)|Z,A=0,X\}&=\int h_0(w,X)dF(w|Z,A=0,X)\label{fred2}
\end{align}
Then under Assumptions \ref{consist}-\ref{complete}, it follows that 
\begin{align}
E(Y|U,A=1,M,X)&=\int h_1(w,M,X)dF(w|U,A=1,M,X)\label{int1}\\
E\{h_1(W,M,X)|U,A=0,X\}&=\int h_0(w,X)dF(w|U,A=0,X)\nonumber %\label{int2}
\end{align}
and furthermore that $E[Y\{1,M(0)\}]$ is identified as 
\begin{align}\label{pmf}
\psi=\int  \int h_0(w,x)dF(w|x)dF(x)
\end{align}
\end{theorem}
The proof of this result, as well as all others in this paper, is given in the Supplementary Material. We name expression (\ref{pmf}) the \textit{proximal mediation formula}, since it generalises Pearl's fundamental mediation formula (\ref{med_form})
%\[\int \int E(Y|A=1,m,l)dF(m|A=0,l)dF(l)\]
to settings where key confounders are unmeasured. Similar to the proximal g-formula in \cite{tchetgen2020introduction}, (\ref{pmf}) is expressed in terms of nested bridge functions. Interestingly, although inferring natural (in)direct effects involves understanding the effect that the exposure has on the mediator, as well as interventions on the mediator, $\psi$ can be identified without a need for additional proxy variables. Nevertheless, compared with proximal identification of the average causal effect, additional restrictions are placed on the exposure- and outcome-inducing proxies, in terms of their relationship to $M$. For example, neither $Z$ nor $W$ are allowed to cause $M$. Such assumptions could be relaxed by collecting data on separate proxies for each bridge function; we will investigate this in future work.
%Nevertheless, whilst such variables are not required for identification, they may be used to strengthen inferences by making for instance the bridge condition (\ref{fred1}) less stringent. Some more general identification results are therefore given in Appendix \ref{extended}. 

Equations (\ref{fred1}) and (\ref{fred2}) refer to inverse problems that are known as Fredholm integral equations of the first kind. We refer to \citet{miao2018identifying} and \citet{cui2020semiparametric} for mathematical conditions that ensure these equations admit solutions. We note that the solutions to these equations are not required to be unique, as all solutions yield a unique value of $\psi$. The identification assumptions may nevertheless be adjusted in order to guarantee a unique solution. Unlike Assumptions \ref{consist} and \ref{lce}-\ref{complete}, the condition that (\ref{fred1}) and (\ref{fred2}) admit solutions is potentially empirically verifiable.

\subsection{Alternative identification strategies}

In this section, we establish two alternative proximal identification results to the proximal mediation formula, which rely on alternative assumptions regarding completeness and the existence of relevant confounding bridge functions, which are given in the Supplementary Material.
\iffalse
\begin{assumption}\label{complete2}(Completeness)
\begin{enumerate}
\item For any function $g(u)$, if $E\{g(U)|W=w,A=1,M=m,X=x\}=0$ for any $w$, $m$ and $x$ almost surely, then $g(U)=0$ almost surely.
\item For any function $g(u)$, if $E\{g(U)|W=w,A=0,X=x\}=0$ for any $w$ and $x$ almost surely, then $g(U)=0$ almost surely. %(CHECK)
\item For any function $g(w)$,  $E\{g(W)|Z=z,A=1,M=m,X=x\}=0$ for any $z$, $m$ and $x$ almost surely, then $g(W)=0$ almost surely.
\item For any function $g(w)$,  $E\{g(W)|Z=z,A=0,X=x\}=0$ for any $z$ and $x$ almost surely, then $g(W)=0$ almost surely. %(CHECK)
\end{enumerate}
\end{assumption}
\fi

\begin{theorem}\label{id_alt}
(Part 1) Suppose that there exists the following confounding bridge functions $h_1(w,M,X)$ that satisfies (\ref{fred1}) and $q_0(z,X)$ that satisfies
\begin{align}
\frac{1}{f(A=0|W,X)}&=E\{q_0(Z,X)|W,A=0,X\}\label{fred3}
\end{align}
Under Assumptions \ref{consist}-\ref{p_proxy} and \ref{complete}.1 above and Assumption \ref{complete2}.2 in the Supplementary Material, we have that
\begin{align}\label{cui_res}
\frac{1}{f(A=0|U,X)}=E\{q_0(Z,X)|U,A=0,X\}
\end{align}
as well as (\ref{int1}), and furthermore that $E[Y\{1,M(0)\}]$ is identified as 
\begin{align*}%\label{pmf2}
\psi=\int  \int I(a=0)q_0(z,x)h_1(w,m,x)dF(w,z,a,m|x)dF(x)
\end{align*}

(Part 2) Alternatively, suppose that there exists confounding bridge functions $q_0(z,X)$ that satisfies (\ref{fred3}) and  $q_1(z,M,X)$ that satisfies
\begin{align}
&E\{q_0(Z,X)|W,A=0,M,X\}\frac{f(A=0|W,M,X)}{f(A=1|W,M,X)}= E\left\{q_1(Z,M,X)|W,A=1,M,X\right\} \label{fred4}
\end{align}
Under Assumptions \ref{consist}-\ref{p_proxy}, and \ref{complete2}.1-\ref{complete2}.2 in the Supplementary Material we have that
\begin{align*}%\label{int4}
&E\{q_0(Z,X)|U,A=0,M,X\}\frac{f(A=0|U,M,X)}{f(A=1|U,M,X)}= E\left\{q_1(Z,M,X)|U,A=1,M,X\right\} %\label{fred4}
\end{align*}
and (\ref{cui_res}), and $E[Y\{1,M(0)\}]$ is identified as 
\begin{align}\label{pmf3}
\psi=\int  \int I(a=1)q_1(z,m,x)ydF(y,z,a,m|x)dF(x)
\end{align}
\end{theorem}
We therefore have three results for proximal identification, each of which relies on two confounding bridge assumptions. The strategy given in the first part of the above theorem relies on a combination of outcome- and treatment-inducing confounding bridge functions, whereas the final strategy relies on two nested treatment-inducing confounding bridge functions. The result (\ref{cui_res}) follows from  \cite{cui2020semiparametric}, and formal conditions for existence of solutions to (\ref{fred3}) and (\ref{fred4}) are given in that paper. %We close this section by noting that the Fredholm equations such as (\ref{fred1}), (\ref{fred2}), (\ref{fred3}) and (\ref{fred4}) are known to be ill-posed, and hence 

%We find it helpful to condition on $M$ in expression (\ref{pmf3}), 
 %$A$ has no causal effect on $W$, conditional on treatment $A$, mediator and measured and unmeasured confounders $X$ and $U$, as well as conditionally on $A$, $X$ and $U$ alone. Also, $Z$ may not causally affect $Y$ of $Z$, 

%and $Y\indep A_M|U,M,L,A_Y$
%More formally, if we replace the previous conditional independences with the assumptions that 
%then 

\section{Semiparametric inference}\label{speff}

\subsection{The semiparametric efficiency bound}

In this section, we will consider inference for $\psi$ under the semiparametric model $\mathcal{M}_{sp}$ which places no restrictions on the observed data distribution besides existence (but not necessarily uniqueness) of bridge functions $h_1$ and $h_0$ that solve (\ref{fred1}) and (\ref{fred2}). Note that assumed existence of the outcome bridge functions places restrictions on the tangent space.  %Technically, these need to exist at all densities that belong to the model. 
Under additional regularity conditions (described below), we can also obtain the semiparametric efficiency bound under $\mathcal{M}_{sp}$.

\begin{assumption}\label{bound}(Regularity conditions):
\begin{enumerate}
\item Let $T_1 :L_2(W,M,X)\to L_2(Z,A=1,M,X)$ denote the operator given by\\ $T_1(g)\equiv E\{g(W,M,X)|Z,A=1,M,X\}$. At the true data generating mechanism, $T_1$ is surjective.
\item Let $T_0 :L_2(W,M,X)\to L_2(Z,A=0,X)$ denote the operator given by\\ $T_0(g)\equiv E\{g(W,M,X)|Z,A=0,X\}$. Then at the true data generating mechanism, $T_0$ is surjective.
\end{enumerate}
\end{assumption}
%QUESTIONS FOR ERIC
%Do we need boundedness condition on exposure models?
%Why do we only need bits involving \epsilon_1 in the closure of the range space.
As noted in \citet{ying2021proximal}, this condition relies on the functions $L_2(W,M,X)$ being rich enough such that any element in $L_2(Z,A=1,M,X)$ and $L_2(Z,A=0,X)$ can be generated via the conditional expectation map. Then we arrive at the following result.

\begin{theorem}\label{eif}
Assume that there exist bridge functions  $h_1$ and $h_0$ at all data laws that belong to the semiparametric model $\mathcal{M}_{sp}$. Furthermore, suppose that at the true data law there exists  $q_0$ and $q_1$ which solve (\ref{fred3}) and (\ref{fred4}), and that Assumption \ref{complete} holds, such that $\psi$ is unique. Then 
\begin{align*}
IF_{\psi}=&I(A=1)q_1(Z,M,X)\{Y-h_1(W,M,X)\}\\&+I(A=0)q_0(Z,X)\{h_1(W,M,X)-h_0(W,X)\}+h_0(W,X)-\psi
\end{align*}
is a valid  influence function for $\psi$ under $\mathcal{M}_{sp}$. %If we further assume that all bridge functions are unique and 
Furthermore, the efficiency bound at the submodel where Assumption \ref{bound} holds and all bridge functions are unique is $E(IF_{\psi}^2)$.
\end{theorem}

\subsection{Multiply robust estimation}

%Ideally, the bridge functions $h_1$, $h_0$, $q_1$ and $q_0$ could be estimated non-parametrically, or using flexible machine learning methods 

%We will touch on inference for $\psi$ based on nonparametric estimation of $h_1$, $h_0$, $q_1$ and $q_0$ %, and the results of Theorem \ref{eif},  
%in the Discussion. For now however, 
We will consider the setting where $L$ is high-dimensional, and parametric working models for $h_1$, $h_0$, $q_0$ and $q_1$ may be useful as a form of dimension-reduction. In that case, we show that an estimator of $\psi$ based on the efficient influence function is multiply robust, in the sense that only certain combinations of these working models need to be correctly specified in order to yield an unbiased estimator. To make this more concrete, consider the following semiparametric models that impose certain restrictions on the observed data distribution:
\begin{align*}
&\mathcal{M}_1: \textrm{$h_1(W,M,X)$ and $h_0(W,X)$ are assumed to be correctly specified};\\
&\mathcal{M}_2: \textrm{$h_1(W,M,X)$ and $q_0(Z,X)$ are assumed to be correctly specified};\\
&\mathcal{M}_3: \textrm{$q_1(Z,M,X)$ and $q_0(Z,X)$ are assumed to be correctly specified}.
\end{align*}
Here, $\mathcal{M}_1$, $\mathcal{M}_2$ and $\mathcal{M}_3$ are all submodels of $\mathcal{M}_{sp}$. The proposed approach will rely on models for the confounding bridge functions, but we will show that only one of $\mathcal{M}_1$, $\mathcal{M}_2$ and $\mathcal{M}_3$ needs to hold to ensure unbiased estimation of the target parameter.

We shall first consider how to obtain inference in the submodels $\mathcal{M}_1$, $\mathcal{M}_2$ and $\mathcal{M}_3$. Let $h_1(W,M,X;\beta_1)$ and $h_0(W,X;\beta_0)$ denote models for the respective bridge functions $h_1(W,M,X)$ and $h_0(W,X)$, indexed by finite-dimensional parameters $\beta_1$ and $\beta_0$.  Likewise, $q_1(Z,M,X;\gamma_1)$ and $q_0(Z,X;\gamma_0)$ are models for the bridge functions $q_1(Z,M,X)$ and $q_0(Z,X)$ indexed by the finite-dimensional parameters $\gamma_1$ and $\gamma_0$ respectively. It follows from  \citet{cui2020semiparametric} that one can obtain estimates $\hat{\beta}_1$, $\hat{\beta}_0$ and $\hat{\gamma}_0$ of $\beta_1$, $\beta_0$ and  $\gamma_0$ as the solutions to the (respective) estimating equations:
\begin{align*}
\sum^n_{i=1}A_i\{Y_i-h_1(W_i,M_i,X_i;\beta_1)\}c_1(Z_i,M_i,X_i)&=0\\
\sum^n_{i=1}(1-A_i)\{h_1(W_i,M_i,X_i;\beta_1)-h_0(W_i,X_i;\beta_0)\}c_0(Z_i,X_i)&=0\\
\sum^n_{i=1}\left\{(1-A_i)q_0(Z_i,X_i;\gamma_0)-1\right\}d_0(W_i,X_i)&=0
\end{align*}
where the first two sets of equations can be solved sequentially; here $c_1(Z_i,M_i,X_i)$ is a function of the same dimension as $\beta_1$, and $c_0(Z_i,X_i)$ and $d_0(W_i,X_i)$ are similarly defined. Although $\hat{\beta}_0$ and hence $h_0(W,X;\hat{\beta}_0)$ depends on $\hat{\beta}_1$, this dependence is suppressed to simplify notation. The above estimating equations can be implemented using software for generalised method of moments or (when models are linear) two-stage least squares. %When $h_1(W,M,X;\beta_1)$ and $h_0(W,X;\beta_0)$ are linear in their parameters, the above estimating equations be straightforwardly implemented using two-stage least squares software.
%It also follows from  \citet{cui2020semiparametric} that an estimate $\hat{\gamma}_0$ of $\gamma_0$ can be obtained as the solution to the equations
%\[\sum^n_{i=1}\left\{(1-A_i)q_0(Z_i,X_i;\gamma_0)-1\right\}d_0(W_i,X_i),\]
%where $d_0(W_i,X_i)$ is of the same dimension as $\gamma_0$. 
Interestingly, despite the fact that (\ref{fred4}) suggests that estimation of $\gamma_0$ would require postulation of a model for $1/f(A=0|W,X)$,  \citet{cui2020semiparametric} show that this is not the case. %Indeed, inferences for $q(Z,X;\gamma_0)$ can be obtained without estimation of the propensity score by using an influence function for $\gamma_0$ as the basis of an estimator (as is done above). 
The efficient choices of $c_1(Z_i,M_i,X_i)$, $c_0(Z_i,X_i)$ and $d_0(W_i,X_i)$ are all implied by results in the Appendix of \citet{cui2020semiparametric}. % ; it is noted there that these efficient choices require modelling of additional components of the observed data law, which may be challenging to model correctly. 
Since the resulting efficiency gain compared to using the choices $c_1(Z_i,M_i,X_i)=(1,Z^T_i,M_i, X^T_i)^T$, $c_0(Z_i,X_i)=(1,Z^T_i, X^T_i)^T$ and $d_0(W_i,X_i)=(1,W^T_i, X^T_i)^T$ is likely to be modest in most situations \citep{stephens2014locally}, we do not consider locally efficient estimation of the nuisance parameters any further. 

Since $q_1$ involves solving an integral equation (\ref{fred4}) involving the ratio of propensity score functions, the results from previous work do not extend to inference for $\gamma_1$. The following theorem then suggests how to obtain semiparametric inference under model $\mathcal{M}_3$, and is more generally relevant for estimation of the average treatment effect in the (un)treated:

%The following theorem therefore 
\begin{theorem}\label{res_if_q1}
All influence functions of regular and asymptotically linear estimators of $\gamma_1$ under the semiparametric model $\mathcal{M}_3$ are of the form
\begin{align*}
-V^{-1}\bigg[&\{Aq_1(Z,M,X;\gamma_1)-(1-A)q_0(Z,X;\gamma_0)\}d_1(W,M,X)\\&-E\left\{\frac{\partial q_0(Z,X;\gamma_0)}{\partial \gamma_0}(1-A)d_1(W,M,X)\right\}\varphi(W,Z,A,X;\gamma_0)\bigg]
\end{align*}
for some function $d_1(W,M,X)$ that is the same dimension as $\gamma_1$, where
\begin{align*}
V=E\left\{\frac{\partial q_1(Z,M,X;\gamma_1)}{\partial \gamma_1}Ad_1(W,M,X)\right\}
\end{align*}
and $\varphi(W,Z,A,X;\gamma_0)$ is the influence function for an estimator of $\gamma_0$. 
\end{theorem}
This theorem indicates that inference for $\gamma_1$ can be obtained without the need to model either $f(A=0|W,M,X)$ or $f(A=1|W,M,X)$ (or their ratio). Indeed, it suggests an estimation strategy for $\gamma_1$; namely, after obtaining $\hat{\gamma}_0$ as previously described, one can obtain $\hat{\gamma}_1$ as the solution to the equations:
\begin{align*}
\sum^n_{i=1}\{A_iq_1(Z_i,M_i,X_i;\gamma_1)-(1-A_i)q_0(Z_i,X_i;\hat{\gamma}_0)\}(1,W^T_i,M_i,X^T_i)^T=0
\end{align*}
Consistent estimation of $\gamma_1$ nevertheless relies on consistent estimation of $\gamma_0$. Given that $q_1(Z,M,X;\gamma_1)$ and $q_0(Z,X;\gamma_0)$ are both confounding bridges for the treatment assignment mechanism, this raises the question of how to postulate models for the two bridge functions that are compatible. A brief discussion on model compatibility comes in the Section \ref{sims}, with more detailed results in the Supplementary Material.

\iffalse
Consistent estimation of $\gamma_1$ nevertheless relies on consistent estimation of $\gamma_0$. Given that $q_1(Z,M,X;\gamma_1)$ and $q_0(Z,X;\gamma_0)$ are both confounding bridge functions  for the treatment assignment mechanism, this raises the question of how to postulate models for the two bridge functions that are compatible. In Appendix \ref{mod_comp}, we show that when $A$ follows a Bernoulli distribution given $U$ and $X$, $M$ is normally distributed given $U$, $A$ and $X$, and $Z\sim \mathcal{N}(\epsilon_0+\epsilon_u U+\epsilon_a A+ \epsilon_x X,\sigma^2_{z|u,a,x})$, then the choices of bridge functions
\begin{align*}
q_0(Z,X)&=1+\exp\{-(\gamma_{0,0}+\gamma_{0,z}Z+\gamma_{0,x}X)\}\\
q_1(Z,M,X)&=\exp(\gamma_{1,0}+\gamma_{0,z}Z+\gamma_{0,z}M+\gamma_{0,x}X)\\
&\quad+\{q_0(Z,X)-1\}\exp\{\gamma_{1,0}+\gamma_{0,0}\epsilon_a+\gamma_{0,z}(Z+\gamma_{0,0}\sigma^2_{z|u,a,x}/2)+\gamma_{0,z}M+\gamma_{0,x}X\}
\end{align*}
satisfy (\ref{fred3}) and (\ref{fred4}). 
\fi

Once we have strategies for estimating nuisance parameters indexing the bridge functions, one can construct proximal outcome regression (P-OR), hybrid (P-hybrid) and IPW (P-IPW) estimators of $\psi$:
\begin{align*}
\hat{\psi}_{P-OR}&=\frac{1}{n}\sum^n_{i=1}h_0(W_i,X_i;\hat{\gamma}_0)\\
\hat{\psi}_{P-hybrid}&=\frac{1}{n}\sum^n_{i=1}(1-A_i)q_0(Z_i,X_i;\hat{\gamma}_0)h_1(W_i,M_i,X_i;\hat{\beta}_1)\\
\hat{\psi}_{P-IPW}&=\frac{1}{n}\sum^n_{i=1}A_iq_1(Z_i,M_i,X_i;\hat{\gamma}_1)Y_i
\end{align*}
Then $\hat{\psi}_{P-OR}$ is a consistent and asymptotically normal (CAN) estimator under model $\mathcal{M}_1$,  $\hat{\psi}_{P-hybrid}$ is CAN under model $\mathcal{M}_2$ and $\hat{\psi}_{P-IPW}$ is CAN under model $\mathcal{M}_3$. Correctly specifying models for the different bridge functions may be challenging, since they are defined as solutions to integral equations, rather than the conditional expectations or probabilities more common in causal inference e.g. $E(Y|A=1,L)$ or $f(A=1|L)$. The development of a proximal multiply robust estimator (P-MR), which enables the relaxation of parametric modelling assumptions, is therefore of interest.
\begin{theorem}\label{res_mr}
Under typical regularity conditions, 
\begin{align*}
\hat{\psi}_{P-MR}=&\frac{1}{n}\sum^n_{i=1}A_iq_1(Z_i,M_i,X_i;\hat{\gamma}_1)\{Y_i-h_1(W_i,M_i,X_i;\hat{\beta}_1)\}\\&+(1-A_i)q_0(Z_i,X_i;\hat{\gamma}_0)\{h_1(W_i,M_i,X_i;\hat{\beta}_1)-h_0(W_i,X_i;\hat{\gamma}_0)\}+h_0(W_i,X_i;\hat{\gamma}_0)
\end{align*}
is a CAN estimator of $\psi$ under the union model $\mathcal{M}_{union}=\mathcal{M}_1\cup \mathcal{M}_2 \cup \mathcal{M}_3$. Furthermore, under model $\mathcal{M}_{union}$, $\hat{\psi}_{P-MR}$ attains the semiparametric efficiency bound at the intersection submodel $\mathcal{M}_1\cap \mathcal{M}_2 \cap \mathcal{M}_3$ where Assumption \ref{bound} also holds.
\end{theorem}
Using standard M-estimation arguments, and the influence functions for the nuisance parameter estimators, one can construct a nonparametric `sandwich' estimator of the standard error for $\hat{\psi}_{P-MR}$ which is robust to potential model misspecification; an alternative option is the nonparametric bootstrap. A weakness of our estimator is that when $Y$ is binary, $\hat{\psi}_{P-MR}$ is not guaranteed to fall within the (0,1) interval. This is an important topic for future work and  could be remedied e.g. by adapting in the proposal in Section 5 of \citet{tchetgen2012semiparametric}.

%and the efficient choice of $c_1(Z_i,M_i,X_i)$  is given in the appendix of \citet{cui2020semiparametric} ($c_0(Z_i,X_i)$ is similarly defined). It also follows from that paper that 

\section{Simulation studies}\label{sims}

In order to evaluate the finite sample performance of the proposed estimators, we conducted a simulation study. Specifically, we generated data ($Y,A,M,X,W,Z$) by $X, U\sim MVN((0.25,0.25,0)^T,\Sigma)$; $f(A=1|X,U)=expit(-(0.5,0.5)^TX-0.4U)$; 
$Z|A,X,U\sim \mathcal{N}(0.2-0.52A+(0.2,0.2)^TX-U,1)$; $W|X,U\sim \mathcal{N}(0.3+(0.2,0.2)^TX-0.6U,1)$ and $M|A,X,U\sim \mathcal{N}(-0.3A-(0.5,0.5)^TX+0.4U,1)$,
where $X=(X_1,X_2)$ and 
\begin{equation*}
\Sigma = 
\begin{pmatrix}
\sigma_{x_1}^2 & \sigma_{x_1x_2} & \sigma_{x_1u} \\
\sigma_{x_1x_2}  & \sigma_{x_2}^2 & \sigma_{x_2u} \\
\sigma_{x_1u} & \sigma_{x_2u} & \sigma_{u}^2  
\end{pmatrix} =
\begin{pmatrix}
0.25 & 0 & 0.05 \\
0 & 0.25 & 0.05 \\
0.05 & 0.05 & 1
\end{pmatrix}.
\end{equation*}
Finally, $Y=2+2A+M+2W-(1,1)^TX-U+2\epsilon^*$ where $\epsilon^* \sim \mathcal{N}(0,1)$. Since $W$ and $M$ are linear in $X$, $U$ and the exposure, it follows that this data generating mechanism is compatible with the following models for $h_1$, $h_0$:
\begin{align*}
h_1(W,M,X;\beta_1)&=\beta_{1,0}+\beta_{1,w}W+\beta_{1,x}^TX+\beta_{1,m}M\\
h_0(W,X;\beta_0)&=\beta_{0,0}+\beta_{0,w}W+\beta_{0,x}^TX
\end{align*}
Furthermore, in the Supplementary Material, we also show that when $A$ follows a Bernoulli distribution given $U$ and $X$, $M$ is normally distributed given $U$, $A$ and $X$ and $Z\sim \mathcal{N}(\epsilon_0+\epsilon_u U+\epsilon_a A+ \epsilon_x X,\sigma^2_{z|u,a,x})$, then the choice of bridge functions
\begin{align*}
q_0(Z,X)&=1+\exp\{-(\gamma_{0,0}+\gamma_{0,z}Z+\gamma_{0,x}X)\}\\
q_1(Z,M,X)&=\exp(\gamma_{1,0}+\gamma_{1,z}Z+\gamma_{1,m}M+\gamma_{1,x}X)\\
&\quad+\{q_0(Z,X)-1\}\exp\{\gamma_{1,0}+\gamma_{0,z}\epsilon_a+\gamma_{1,z}(Z+\gamma_{0,z}\sigma^2_{z|u,a,x})+\gamma_{1,m}M+\gamma_{1,x}X\}
\end{align*}
satisfies (\ref{fred3}) and (\ref{fred4}). Under the additional constraint that $\epsilon_a=-\gamma_{1,z}\sigma^2_{z|u,a,x}$ (which is enforced here), it follows that the expression for $q_1$ simplifies to 
\begin{align*}
q_1(Z,M,X)&=q_0(Z,X)\exp(\gamma_{1,0}+\gamma_{1,z}Z+\gamma_{1,z}M+\gamma_{1,x}X).
\end{align*}

Let $\hat{\delta}_{P-DR}$ be the proximal doubly robust (P-DR) estimator of $E\{Y(0)\}$ considered in \citet{cui2020semiparametric}. We considered four proximal mediation estimators of the natural direct effect $E\{Y(1,M(0))\}-E\{Y(0)\}$:
$\hat{\theta}_{P-OR}=\hat{\psi}_{P-OR}-\hat{\delta}_{P-DR}$, 
$\hat{\theta}_{P-hybrid}=\hat{\psi}_{P-hybrid}-\hat{\delta}_{P-DR}$,
$\hat{\theta}_{P-IPW}=\hat{\psi}_{P-IPW}-\hat{\delta}_{P-DR}$ and $\hat{\theta}_{P-MR}=\hat{\psi}_{P-MR}-\hat{\delta}_{P-DR}$. %We also considered a na\"ive non-proximal estimator $\hat{\theta}_{OLS}$ of the direct effect, based on linearly regressing $Y$ on $A$, $M$ and $L$.
We evaluated the performance of the proposed settings in settings where either all bridge functions were correctly modelled (Experiment 1), $q_1$ and $q_0$ were misspecified (Experiment 2), $q_1$ and $h_0$ were misspecified (Experiment 3), or $h_1$ and $h_0$ were misspecified (Experiment 4). We misspecified the models by including $|X_1|^{1/2}$ and $|X_2|^{1/2}$ in the bridge function rather than $X=(X_1,X_2)^T$. We conducted simulations at $n=2,000$ and repeated each experiment $1,000$ times.

\begin{table}[htbp]
\caption{Simulation results from Experiments 1-4. Exp: Experiment; Est: estimator; MSE: mean squared error; Bias: Monte Carlo bias; Coverage: 95\% confidence interval (CI) coverage; Mean Length: average 95\% CI length; Med. Length: median 95\% CI length.}
\centering
\begin{tabular}{lllllll}
  \hline
Exp & Est & Bias & MSE & Coverage & Mean Length & Med. Length \\ 
  \hline
1 & $\hat{\theta}_{P-IPW}$ & 0.00 & 0.02 & 0.95 & 0.51 & 0.50 \\ 
   & $\hat{\theta}_{P-hybrid}$ & 0.00 & 0.02 & 0.96 & 0.50 & 0.50 \\ 
   & $\hat{\theta}_{P-OR}$ & 0.00 & 0.02 & 0.96 & 0.50 & 0.50 \\ 
   & $\hat{\theta}_{P-MR}$ & 0.00 & 0.02 & 0.95 & 0.51 & 0.50 \\ 
  2 & $\hat{\theta}_{P-IPW}$ & 0.19 & 0.05 & 0.71 & 0.52 & 0.51 \\ 
   & $\hat{\theta}_{P-hybrid}$ & -0.15 & 0.04 & 0.82 & 0.52 & 0.52 \\ 
   & $\hat{\theta}_{P-OR}$ & 0.00 & 0.02 & 0.95 & 0.50 & 0.50 \\ 
   & $\hat{\theta}_{P-MR}$ & 0.00 & 0.02 & 0.95 & 0.50 & 0.50 \\ 
  3 & $\hat{\theta}_{P-IPW}$ & 0.38 & 0.16 & 0.30 & 0.60 & 0.59 \\ 
   & $\hat{\theta}_{P-hybrid}$ & 0.00 & 0.02 & 0.95 & 0.50 & 0.50 \\ 
   & $\hat{\theta}_{P-OR}$ & -0.14 & 0.04 & 0.81 & 0.52 & 0.51 \\ 
   & $\hat{\theta}_{P-MR}$ & 0.00 & 0.02 & 0.95 & 0.51 & 0.50 \\ 
  4 & $\hat{\theta}_{P-IPW}$ & -0.01 & 0.02 & 0.94 & 0.51 & 0.51 \\ 
   & $\hat{\theta}_{P-hybrid}$ & 0.21 & 0.06 & 0.66 & 0.53 & 0.53 \\ 
   & $\hat{\theta}_{P-OR}$ & 0.17 & 0.05 & 0.72 & 0.52 & 0.51 \\ 
   & $\hat{\theta}_{P-MR}$ & -0.01 & 0.02 & 0.95 & 0.51 & 0.51 \\ 
   \hline
   \end{tabular}
\label{result_tab}
\end{table}

The results are given in Table \ref{result_tab}. As a benchmark, we also considered a na\"ive non-proximal estimator $\hat{\theta}_{OLS}$ of the direct effect, based on linearly regressing $Y$ on $A$, $M$ and $L$; its Monte Carlo bias was 0.5, with 95\% confidence intervals that included the true value only 29\% of the time. When all bridge functions were correctly specified, the different proximal estimators had comparable performance, with $\hat{\theta}_{P-OR}$ and $\hat{\theta}_{P-MR}$ exhibiting slightly greater efficiency compared with the other methods. Across the different mechanisms of misspecification considered, we see that only the multiply robust estimator continues to have low bias, with confidence intervals that possess (approximately) their advertised coverage. %In experiments where $h_1$ and/or $h_0$ were misspecified, in certain simulations the intervals for $\hat{\theta}_{P-OR}$ and $\hat{\theta}_{P-MR}$ could become wide, as reflected in the discrepancy between the mean and median interval length. 

%  y<-2+2*a+m-l%*%c(1,1)+2*w-u-0.5*z+2*rnorm(n)
%  w<-rnorm(n,0.3+l%*%c(0.2,0.2)-0.6*u+0.2*a)
In a further set of simulation studies, we considered sensitivity of the different proposals to changes in the confounding mechanism as well as violations of the structural assumptions that underpin the methods. First, in Experiment 5 we changed the above data-generating mechanism such that $f(A=1|X,U)$, $f(M|A,X,U)$ and $f(Y|A,M,W,X,U)$ no longer depended on $U$, to check how the methods performed when there was no unmeasured confounding. In this specific setting, to construct the benchmark estimator $\hat{\theta}_{OLS}$ we adjusted for $A$, $M$, $X$ and $W$ \textit{but not} $Z$, to avoid collider bias induced via an association between $A$ and $U$. In Experiment 6, we considered violations of the exclusion restriction Assumption \ref{p_proxy}.2, by generating $Y= 2+2A+M+2W-(1,1)^TX-U-0.5Z+2\epsilon^*$. In Experiment 7, we considered violations of the exclusion restriction Assumption \ref{p_proxy}.4, by generating $W$ from $W|X,U,A\sim \mathcal{N}(0.3+(0.2,0.2)^TX-0.6U+0.2A,1)$. In Experiment 8, we looked at a near-violation of the completeness conditions, where the coefficient for $U$ in the model for $E(W|X,U)$ was reduced in strength to $0.05$, such that $W$ was only weakly $U$- relevant and the conditional associations between $W$ and $Z$ were also weak. In 9), we reversed this such that $Z$ was now only weakly $U$-relevant.
%the completeness assumptions were violated since $f(W|X,U)$ no longer depended on $U$, 
%; in 9), $f(Z|A,X,U)$ no longer depended on $U$. 
Each of the models used to construct the bridge functions included $X$, rather than the transformations considered in Experiments 2-4. Values of the parameters in the data-generating mechanism were adjusted where necessary, to ensure that all bridge functions were correctly specified. The results of Experiments 5-9 are shown in Table\ref{results_tab2} in the Supplementary Material. We see that in settings where conditional exchangeability holds, the proximal estimators perform similarly in terms of bias compared with $\hat{\theta}_{OLS}$, but displayed a decrease in efficiency. When the exclusion restrictions are violated, in this case the proximal estimators performed similarly or slightly worse than the na\"ive, biased OLS estimator. When $Z$ or $W$ are not $U$-relevant, we see that the standard errors for the proximal estimators can dramatically inflate; this is unsurprising, given how common instrumental variable estimators perform when instruments are weak/irrelevant. The proximal estimators also typically displayed considerably larger bias than $\hat{\theta}_{OLS}$, but smaller median bias. The multiply robust estimator $\hat{\theta}_{P-MR}$ generally displayed smaller mean and median bias compared with the other methods in Experiments 8 and 9. 

\section{Data analysis}

In the Job Corps study, participants were randomised from November 1994 to February 1996 either to treatment (access to the Job Corps program) or to control (no access).  However, since individuals could choose whether to participate in the program or not, we will treat the exposure of interest (\textit{did the individual attend class in the year following assignment?}) as non-randomised. The outcome of interest was the number of arrests in the fourth year after assignment, and the mediator of interest was the percentage of weeks employed in the second year. Although data on a rich set of covariates was measured at baseline, it was nevertheless possible that unmeasured confounding could lead to biased estimates of both direct and indirect effects. Investigators collected information on factors that were known to be associated with duration in the program e.g. expectations of the program, interactions with recruiters. \citet{huber2020direct} note that such variables may be strongly correlated with motivation, considered as an important latent source of confounding. They therefore adjusted for these variables in the analysis in the same way as standard measured confounders. In contrast, we treated these variables as proxies of a unmeasured confounder, and changed the analysis accordingly. Similar to \citet{tchetgen2020introduction}, we restricted consideration to four potential proxies strongly correlated with the expose and/or the outcome: time being spent spoken to by recruiter, worried about the Job Corps program, expected improvement in social skills and whether the first contact by the recruiter was in the office or not. Since there was no \textit{a priori} understanding as to whether these were candidates for $Z$ or $W$, we used the algorithm described in \citet{tchetgen2020introduction} to assign them. This led to the first two being used for $Z$, and the second two used for $W$.

Our sample consisted of 10,775 participants; further information on the sample is given in \citet{huber2020direct}. 257 participants had missing data on $Z$, who were removed from the analysis. Linear and logistic models were postulated for the outcome and exposure bridge functions respectively. We considered both proximal IPW, hybrid, outcome regression and multiply robust estimators of $E[Y\{1,M(0)\}]$ and $E[Y\{0,M(1)\}]$. We contrasted our estimator with a standard multiply robust (S-MR) approach $\hat{\theta}_{S-MR}$ based on the efficient influence function derived in 
\citet{tchetgen2012semiparametric} that is valid under conditional exchangeability-type assumptions; we again postulated linear models for $E(Y|A=a,M,X,Z,W)$ and $E\{E(Y|A=a,M,X,Z,W)|A=a^*,X,Z,W\}$ and a logistic model for the odds $P(A=1|M,X,Z,W)/P(A=0|M,X,Z,W)$ and for $P(A=1|X,Z,W)$. All outcome regression models for the confounding bridge functions and otherwise were fit separately in control and treatment groups, to allow for treatment-mediator and treatment-covariate interactions. Since the set of covariates measured at baseline was relatively high in dimension,
we excluded variables with amounts of missingness $>$50\%, or which had some missing values and were highly correlated with variables that were fully observed. For all other variables with missingness, we used the missing indicator method as in \citet{huber2020direct}.  Standard errors for all estimators were calculated using sandwich estimators.
In the Supplementary Material, following the AGReMA statement on good practice  for conducting and reporting mediation analysis \citep{lee2021guideline}, we provide further information on the study and data analysis.

\begin{table}[ht]
\caption{Results from the analysis of the Job Corps study. CI: confidence interval.}
\label{jobs_tab_main}
\centering
\begin{tabular}{rrrrrrr}
  \hline
Estimand & S-MR & 95\% CI  & P-MR & 95\% CI \\ 
  \hline
$E[Y\{1,M(0)\}]-[Y\{0,M(0)\}]$ & -0.0107 & -0.0341, 0.0128 & 0.0057 & -0.0699, 0.0813  \\ 
$E[Y\{1,M(1)\}]-[Y\{0,M(1)\}]$ & -0.0110 & -0.0345, 0.0124 & -0.0016 & -0.0781, 0.0749 \\ 
$E[Y\{1,M(1)\}]-[Y\{1,M(0)\}]$ & 0.0006 & -0.0229, 0.0241 & -0.0088 & -0.0314, 0.0138 \\ 
$E[Y\{0,M(1)\}]-[Y\{0,M(0)\}]$ & 0.0009 & -0.0001, 0.0019 & -0.0016 & -0.0110, 0.0078\\ 
   \hline
\end{tabular}
\end{table}

Results for the standard and proximal multiply robust approaches can be seen in Table \ref{jobs_tab_main}. The total effect estimate given by the standard approach was -0.01 (95\% confidence interval (CI): -0.022, 0.002) and for the proximal approach, it was -0.003 (95\% CI: -0.040, 0.033). The estimates of the direct and indirect effects yielded by both approaches are also close to the null, and all 95\% confidence intervals contain the null. The direct effects estimates under the proximal approach tended to be closer to the null; and the indirect effects were slightly larger in magnitude (although still very small). The results of the other estimators can be found in the Supplementary Material; the multiply robust estimator $\hat{\theta}_{P-MR}$ tended to agree more closely with   $\hat{\theta}_{P-hybrid}$, although there was not a large disparity between the point estimates.

%Direct Effect 0   -0.0089146301 -0.0322202892 0.014391029
%Indirect Effect 1 -0.0092336182 -0.0325351726 0.014067936
%Direct Effect 1    0.0003789911 -0.0006189531 0.001376935
%Indirect Effect 0  0.0006979792 -0.0001451750 0.001541133

%Direct Effect 0   -0.04293060 -0.16877206 0.08291085
%Indirect Effect 1 -0.03118134 -0.13789851 0.07553584
%Direct Effect 1    0.01363373 -0.08123621 0.10850367
%Indirect Effect 0  0.00188446 -0.02336791 0.02713683

%Results can be seen in Table \ref{jobs_tab}. The proximal and non-proximal analyses broadly agree in this case, indicating a direct effect of treatment on depression symptoms. %This suggests that the original analyses may not be prone to bias due to unmeasured confounding. 

%It hence offers some reassurance, particularly in view of the previous sensitivity analysis. The direct effect with $M$ taken at the level under control was slightly more pronounced than the direct effect at $M(1)$; as in \citet{imai2010general}, none of the estimates were statistically significant at the 5\% level. We see that $\hat{\theta}_{S-MR}$, $\hat{\theta}_{P-hybrid}$ and $\hat{\theta}_{P-MR}$ are close to each other, whilst  $\hat{\theta}_{P-IPW}$ is further away, indicating that the model for $q_1$ may be misspecified.

\section{Discussion}\label{disc}

%In this work, we have provided three proximal identification results for natural direct and indirect effects, and have also derived the efficient influence function for the mediation functional under a nonparametric model. The efficient influence function involves four separate confounding bridge functions. When parametric working models are used for these confounding bridge functionals, the resulting estimators of natural (in)direct effects are multiply robust. %Specifically, our estimators are unbiased so long as both outcome confounding bridge models are correct, or two exposure confounding bridge models are correct, or a specific combination of outcome and confounding bridge models are both correct. 

An advantage of doubly/multiply robust methods, used in combination with cross-fitting, is that data-adaptive methods can be used to estimate nuisance parameters, yet their potentially slow rates of convergence are not necessarily inherited by the estimator of the target parameter \citep{chernozhukov2018double}. A complication in proximal learning is that the nuisances are defined as the solutions to integral equations. Progress in this direction is described in \citet{ghassami2022minimax} and \citet{kallus2021causal}; it would thus be useful to extend these ideas to mediation analysis. Another avenue for future work would be to extend the results of identification and estimation to more general path-specific effects  \citep{avin2005identifiability,shpitser2013counterfactual}, which are relevant in particular in settings when confounders of the mediator-outcome relationship are affected by the exposure. In such cases the cross-world assumption fails to hold, and standard natural effects are no longer identified. Finally, an important topic more generally in proximal learning is the development of sensitivity analysis methods. A simple way to check how sensitive results are to categorisation of the $Z$ and $W$ proxies is to permute the labels. The development of more advanced tools to assess deviations from specific key assumptions (e.g. the exclusion restrictions involving the proxies) is left to future work. Under the failure of certain assumptions, methods for \textit{partial identification} such as  nonparametric bounds may also be useful \citep{robins1989analysis,manski1990nonparametric}.

%In particular, tools to explore the sensitivity of results to violations of the key excluding restrictions involving the proxies could help encourage the uptake of this framework. 
%Next theorem, triple robustness. 
%Will want to note motivate two models for the exposure.
%Show variation independence of q model.

\section*{Acknowledgements}

The first author gratefully acknowledges support from the Ghent University Special Research Fund and the Research Foundation Flanders. The second and third authors gratefully acknowledge support from the National Institutes of Health.

\bibliographystyle{biometrika}
\bibliography{proxy_mediation_biblio}

\appendix

\section*{Appendix}

\subsection*{Additional figures}\label{add_fig}

\begin{figure}[H]
\centering
\subfigure[]{%
 	\begin{tikzpicture}[scale=0.8]
%		\node[] (r) at (0, 4)   {$(b)$};
	\node[] (a) at (0, 2)   {$A$};
	\node[] (m) at (3.5, 4)   {$M$};
	\node[] (y) at (7, 2)   {$Y$};
	%\node[]  (x) at (3.5,0) 	 {$X$};
	\node[draw,circle]  (u) at (3.5,0) 	 {$U$};
	\node[] (w) at (7,0) 	 {$W$};
	\node[] (z) at (0,0) 	 {$Z$};
	%\node[] (z) at (4,1.25) 	 {$Z$};
	%\node[] (v) at (0,4) 	 {$V$};

	\path[->] (a) edge node {} (y);
	\path[->] (a) edge node {} (m);
	\path[->] (m) edge node {} (y);
	\path[->] (u) edge node {} (a);
	\path[->] (u) edge node {} (z);
	%\path[->] (x) edge node {} (a);
	%\path[->] (x) edge node {} (m);
	%\path[->] (x) edge node {} (y);
	%\path[->] (x) edge node {} (z);
	%\path[->] (x) edge node {} (w);
	\path[->] (z) edge node {} (a);
	\path[->] (u) edge node {} (y);
	\path[->] (u) edge node {} (w);
	\path[->] (u) edge node {} (m);
	%\path[->] (u) edge node {} (z);
	%\path[->] (z) edge node {} (m);
	%\path[->] (u) edge [bend right=45] (m);
	\path[->] (w) edge node {} (y);
	%\path[->] (a) edge node {} (v);
	%\path[->] (u) edge node {} (v);
	%\path[->] (v) edge node {} (m);
	\end{tikzpicture}
  \label{fig:subfigure1}}
\quad
\subfigure[]{%
 	\begin{tikzpicture}[scale=0.8]
%		\node[] (r) at (0, 4)   {$(b)$};
	\node[] (a) at (0, 2)   {$A$};
	\node[] (m) at (3.5, 4)   {$M$};
	\node[] (y) at (7, 2)   {$Y$};
	\node[draw,circle]  (u) at (3.5,0) 	 {$U$};
	\node[] (w) at (7,0) 	 {$W$};
	\node[] (z) at (0,0) 	 {$Z$};

	\path[->] (a) edge node {} (y);
	\path[->] (a) edge node {} (m);
	\path[->] (m) edge node {} (y);
	\path[->] (u) edge node {} (a);
	%\path[->] (u) edge node {} (z);
	%\path[->] (x) edge node {} (a);
	%\path[->] (x) edge node {} (m);
	%\path[->] (x) edge node {} (y);
	%\path[->] (x) edge node {} (z);
	%\path[->] (x) edge node {} (w);
	\path[->] (z) edge node {} (a);
	\path[->] (u) edge node {} (y);
	\path[->] (u) edge node {} (w);
	\path[->] (u) edge node {} (m);
	%\path[->] (u) edge node {} (z);
	%\path[->] (z) edge node {} (m);
	%\path[->] (u) edge [bend right=45] (m);
	\path[->] (w) edge node {} (y);
	%\path[->] (a) edge node {} (v);
	%\path[->] (u) edge node {} (v);
	%\path[->] (v) edge node {} (m);
	\end{tikzpicture}
  \label{fig:subfigure2}}
\subfigure[]{%
 	\begin{tikzpicture}[scale=0.8]
%		\node[] (r) at (0, 4)   {$(b)$};
	\node[] (a) at (0, 2)   {$A$};
	\node[] (m) at (3.5, 4)   {$M$};
	\node[] (y) at (7, 2)   {$Y$};
	\node[draw,circle]  (u) at (3.5,0) 	 {$U$};
	\node[] (w) at (7,0) 	 {$W$};
	\node[] (z) at (0,0) 	 {$Z$};

	\path[->] (a) edge node {} (y);
	\path[->] (a) edge node {} (m);
	\path[->] (m) edge node {} (y);
	\path[->] (u) edge node {} (a);
	%\path[->] (u) edge node {} (z);
	%\path[->] (x) edge node {} (a);
	%\path[->] (x) edge node {} (m);
	%\path[->] (x) edge node {} (y);
	%\path[->] (x) edge node {} (z);
	%\path[->] (x) edge node {} (w);
	\path[->] (z) edge node {} (a);
	\path[->] (u) edge node {} (y);
	\path[->] (w) edge node {} (u);
	\path[->] (u) edge node {} (m);
	%\path[->] (u) edge node {} (z);
	%\path[->] (z) edge node {} (m);
	%\path[->] (u) edge [bend right=45] (m);
	\path[->] (w) edge node {} (y);
	%\path[->] (a) edge node {} (v);
	%\path[->] (u) edge node {} (v);
	%\path[->] (v) edge node {} (m);
	\end{tikzpicture}
  \label{fig:subfigure3}}
\quad
\subfigure[]{%
 	\begin{tikzpicture}[scale=0.8]
%		\node[] (r) at (0, 4)   {$(b)$};
	\node[] (a) at (0, 2)   {$A$};
	\node[] (m) at (3.5, 4)   {$M$};
	\node[] (y) at (7, 2)   {$Y$};
	\node[draw,circle]  (u) at (3.5,0) 	 {$U$};
	\node[] (w) at (7,0) 	 {$W$};
	\node[] (z) at (0,0) 	 {$Z$};

	\path[->] (a) edge node {} (y);
	\path[->] (a) edge node {} (m);
	\path[->] (m) edge node {} (y);
	\path[->] (u) edge node {} (a);
	%\path[->] (u) edge node {} (z);
	%\path[->] (x) edge node {} (a);
	%\path[->] (x) edge node {} (m);
	%\path[->] (x) edge node {} (y);
	%\path[->] (x) edge node {} (z);
	%\path[->] (x) edge node {} (w);
	\path[->] (a) edge node {} (z);
	\path[->] (u) edge node {} (y);
	\path[->] (u) edge node {} (w);
	\path[->] (u) edge node {} (m);
	\path[->] (u) edge node {} (z);
	%\path[->] (z) edge node {} (m);
	%\path[->] (u) edge [bend right=45] (m);
	\path[->] (w) edge node {} (y);
	%\path[->] (a) edge node {} (v);
	%\path[->] (u) edge node {} (v);
	%\path[->] (v) edge node {} (m);
	\end{tikzpicture}
  \label{fig:subfigure4}}
\subfigure[]{%
 	\begin{tikzpicture}[scale=0.8]
%		\node[] (r) at (0, 4)   {$(b)$};
	\node[] (a) at (0, 2)   {$A$};
	\node[] (m) at (3.5, 4)   {$M$};
	\node[] (y) at (7, 2)   {$Y$};
	\node[draw,circle]  (u) at (3.5,0) 	 {$U$};
	\node[] (w) at (7,0) 	 {$W$};
	\node[] (z) at (0,0) 	 {$Z$};

	\path[->] (a) edge node {} (y);
	\path[->] (a) edge node {} (m);
	\path[->] (m) edge node {} (y);
	\path[->] (u) edge node {} (a);
	%\path[->] (u) edge node {} (z);
	%\path[->] (x) edge node {} (a);
	%\path[->] (x) edge node {} (m);
	%\path[->] (x) edge node {} (y);
	%\path[->] (x) edge node {} (z);
	%\path[->] (x) edge node {} (w);
	\path[->] (a) edge node {} (z);
	\path[->] (u) edge node {} (y);
	\path[->] (w) edge node {} (u);
	\path[->] (u) edge node {} (m);
	\path[->] (u) edge node {} (z);
	%\path[->] (z) edge node {} (m);
	%\path[->] (u) edge [bend right=45] (m);
	\path[->] (w) edge node {} (y);
	%\path[->] (a) edge node {} (v);
	%\path[->] (u) edge node {} (v);
	%\path[->] (v) edge node {} (m);
	\end{tikzpicture}
  \label{fig:subfigure5}}
\quad
\subfigure[]{%
 	\begin{tikzpicture}[scale=0.8]
%		\node[] (r) at (0, 4)   {$(b)$};
	\node[] (a) at (0, 2)   {$A$};
	\node[] (m) at (3.5, 4)   {$M$};
	\node[] (y) at (7, 2)   {$Y$};
	\node[draw,circle]  (u) at (3.5,0) 	 {$U$};
	\node[] (w) at (7,0) 	 {$W$};
	\node[] (z) at (0,0) 	 {$Z$};

	\path[->] (a) edge node {} (y);
	\path[->] (a) edge node {} (m);
	\path[->] (m) edge node {} (y);
	\path[->] (u) edge node {} (a);
	%\path[->] u) edge node {} (z);
	%\path[->] (x) edge node {} (a);
	%\path[->] (x) edge node {} (m);
	%\path[->] (x) edge node {} (y);
	%\path[->] (x) edge node {} (z);
	%\path[->] (x) edge node {} (w);
	\path[->] (z) edge node {} (a);
	\path[->] (u) edge node {} (y);
	\path[->] (u) edge node {} (w);
	\path[->] (u) edge node {} (m);
	\path[->] (z) edge node {} (u);
	%\path[->] (z) edge node {} (m);
	%\path[->] (u) edge [bend right=45] (m);
	\path[->] (w) edge node {} (y);
	%\path[->] (a) edge node {} (v);
	%\path[->] (u) edge node {} (v);
	%\path[->] (v) edge node {} (m);
	\end{tikzpicture}
  \label{fig:subfigure6}}
\caption{Other causal diagrams compatible with Assumptions \ref{lce} and \ref{p_proxy}}
\label{examples}
\end{figure}
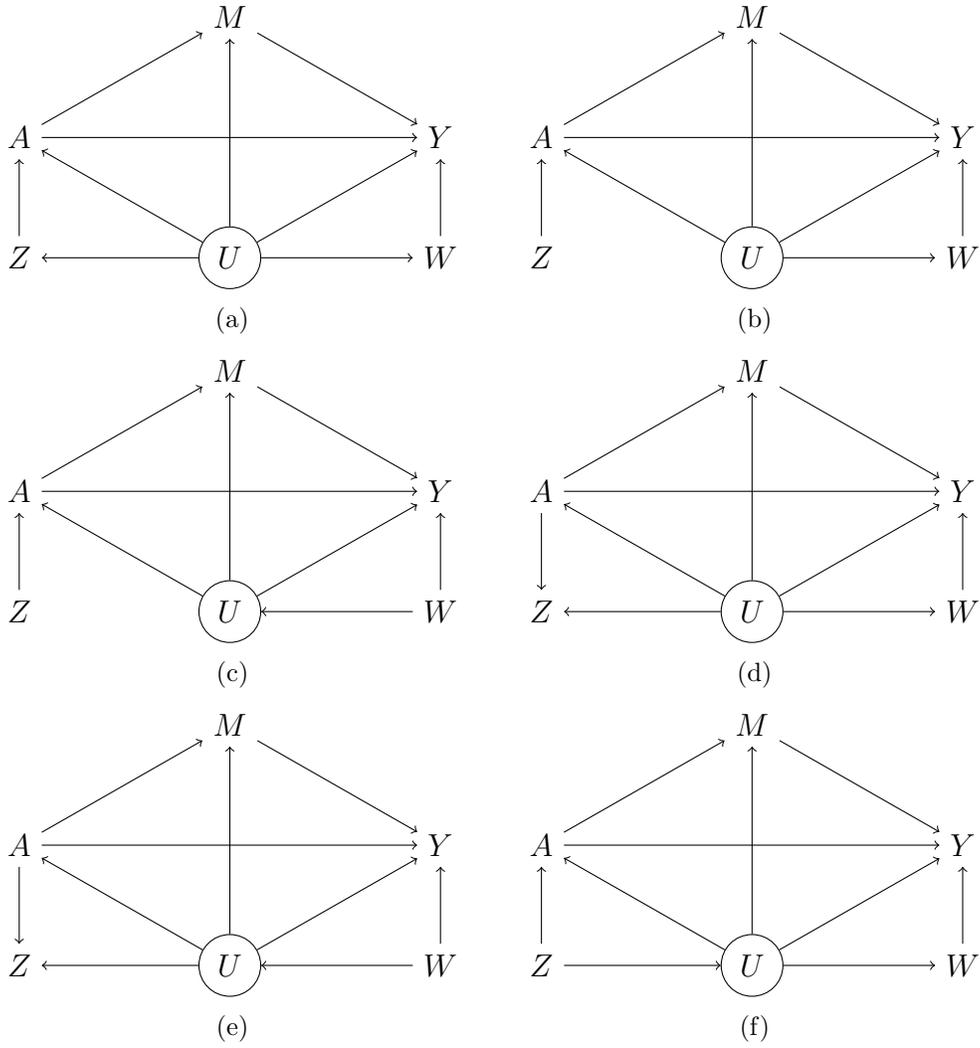

\subsection*{Completeness}

In the present context, the two components of Assumption \ref{complete} demand sufficient variability in $Z$ relative to the variability in $U$, and that $Z$ is associated with $U$ (both conditional on $A=1$, $M$, and $X$, and conditional on $A=0$ and $X$, at any given values of the mediators, covariates and proxy $Z$). %The second two components require that $Z$ has sufficient variation relative to $W$, and that $Z$ is conditionally associated with $W$. %Together, 
They enable one to use the measured proxies to learn about the unmeasured $U$. %; note that only Assumptions \ref{complete}.3 and \ref{complete}.4 can be checked from the observed data. 
Some specific examples regarding completeness are given below:
\begin{example}{(Binary confounder)}
Suppose that $W$, $Z$ and $U$ are all binary. Then Assumption \ref{complete} reduces to the conditions that $W \nindep Z|A=1,M=m,X=x$ for all $m$ and $x$, and that $W \nindep Z|A=0,X=x$ for all $x$.
\end{example}
\begin{example}{(Categorical confounder)}
Suppose now that $W$, $Z$ and $U$ are categorical, (with number of categories $d_u$, $d_z$ and $d_w$ respectively), then Assumption \ref{complete} requires that $\textrm{min}(d_z,d_w)\geq d_u$ or in other words, that $Z$ and $W$ should each have \textit{at least} as many categories as $U$. Furthermore, it also implies a rank conditions on relevant matrices based on the conditional distribution of $W$ given $Z$, $A=1$, $M$ and $X$, as well as $W$ given $Z$, $A=0$ and $X$ (see \cite{shi2020multiply} for details). 
\end{example}
\begin{example}{(Exponential families)}
Suppose that the distribution of $U$ given $Z$, $A=1$, $M$ and $X$ is absolutely continuous with probability approaching one, with density 
\[f(U=u|Z=z,A=1,M=m,X=x)=s(u)t(z,m,x)\exp\{\mu(z,m,x)^T\tau(u)\}\]
where $s(u)>0$, $\tau(u)$ is one-to-one in $u$ and the support of $\mu(z,m,x)$ is an open set. Then Assumption \ref{complete}(i) holds \citep{newey2003instrumental}, and the remaining condition in Assumption \ref{complete} can be shown similarly.
\end{example}
Sufficient conditions for completeness have also been given for location-scale families \citep{hu2018nonparametric}, as well as in  nonparametric models \citep{d2011completeness,darolles2011nonparametric}. One can can see (in particular from the order condition in Example 2) that accumulating data on a rich collection of proxy variables provides a better opportunity to control for an unmeasured $U$. For example, if $Z$ and $W$ are categorical but $U$ has infinite support, the completeness conditions would generally fail to hold.

\subsection*{Derivation of conditional independencies}

In the proofs that follow, we will make repeated use of conditional independencies that are implied by Assumptions \ref{consist}, \ref{lce} and \ref{p_proxy}. We will make use of the graphoid axioms of conditional independence for random variables $R_1$, $R_2$, $R_3$, $R_4$ \citep{dawid1979conditional}:
\begin{enumerate}
    \item $R_1 \indep R_2|R_3 \rightarrow R_2 \indep R_1|R_3$.
    \item $R_1 \indep (R_2,R_4)|R_3 \rightarrow R_1 \indep R_2|R_3$ and $R_1 \indep R_4|R_3$.
    \item $R_1 \indep (R_2,R_4)|R_3 \rightarrow R_1 \indep R_2|R_3,R_4$.
        \item $R_1 \indep R_2|R_3$ and $R_1 \indep R_4|R_3,R_2\rightarrow R_1 \indep (R_2,R_4)|R_3$.
\end{enumerate}
in order to show to show that 
\begin{align*}
&Z \indep Y|M,A,U,X \\
&Z\indep M|A,U,X\\
&Z \indep W|M,A,U,X\\
&Z \indep (W,M)|A,U,X\\
&Z \indep W|A,U,X\\
&W \indep A|U,X\\
&W \indep A|M,U,X
\end{align*}

Then
\iffalse
\begin{align*}
&Z\indep Y(a,m),M(a)|A,U,X \quad \quad \textrm{(Assumptions \ref{p_proxy}.1, \ref{p_proxy}.2 and 4th axiom)}\\
&\to Z\indep Y(m),M|A,U,X \quad \quad \textrm{(Assumption \ref{consist})}\\
&\to Z\indep Y(m)|M,A,U,X \quad \quad \textrm{(3rd axiom)}\\
&\to Z \indep Y|M,A,U,X \quad \quad \textrm{(Assumption \ref{consist})}
\end{align*}
\fi
\begin{align*}
&Y(a,m) \indep (Z,M(a))|A,U,X \quad \quad \textrm{(Assumptions \ref{lce}, \ref{p_proxy}.1, and 4th axiom)}\\
&\to Y(m)\indep (Z,M)|A,U,X \quad \quad \textrm{(Assumption \ref{consist})}\\
&\to Y(m)\indep Z|M,A,U,X \quad \quad \textrm{(3rd axiom)}\\
&\to Y \indep Z|M,A,U,X \quad \quad \textrm{(Assumption \ref{consist})}\\
&\to Z \indep Y|M,A,U,X \quad \quad \textrm{(1st axiom)}
\end{align*}

and 
\begin{align*}
&Z\indep M(a),A|U,X \quad \quad \textrm{(Assumptions \ref{lce}.3, \ref{p_proxy}.2 and 4th axiom)}\\
&\to Z\indep M(a)|A,U,X \quad \quad \textrm{(3rd axiom)}\\
&\to Z\indep M|A,U,X \quad \quad \textrm{(Assumption \ref{consist})}
\end{align*}
and
\begin{align*}
&W \indep (M(a),A,Z)|U,X \quad \quad \textrm{(Assumptions \ref{p_proxy}.3, \ref{p_proxy}.4 and 4th axiom)}\\
&\to W \indep (M(a),Z)|A,U,X \quad \quad \textrm{(3rd axiom)}\\
& \to W \indep (M,Z)|A,U,X \quad \quad \textrm{(Assumption \ref{consist})}\\
& \to W \indep Z|M,A,U,X \quad \quad \textrm{(3rd axiom)}\\
& \to Z \indep W|M,A,U,X \quad \quad \textrm{(1st axiom)}
\end{align*}
From the previous two results and the 4th axiom, we have $Z \indep (W,M)|A,U,X$. Also,
\begin{align*}
&W \indep (M(a),A,Z)|U,X \quad \quad \textrm{(Assumptions \ref{p_proxy}.3, \ref{p_proxy}.4 and 4th axiom)}\\
&\to W \indep (A,Z)|U,X \quad \quad \textrm{(2nd axiom)}\\
& \to W \indep Z|A,U,X \quad \quad \textrm{(3rd axiom)}
\end{align*}

%and combining the last two results gives
%\[Z\indep W,M|A,U,X\]
Next,
\begin{align*}
&W \indep (M(a),A,Z)|U,X \quad \quad \textrm{(Assumptions \ref{p_proxy}.3, \ref{p_proxy}.4 and 4th axiom)}\\
&\to W \indep A|U,X \quad \quad \textrm{(2nd axiom)}
\end{align*}
Finally,
\begin{align*}
&W \indep (M(a),A,Z)|U,X \quad \quad \textrm{(Assumptions \ref{p_proxy}.3, \ref{p_proxy}.4 and 4th axiom)}\\
&\to W \indep (M(a),A)|U,X \quad \quad \textrm{(2nd axiom)}\\
&\to W \indep M(a)|A,U,X \quad \quad \textrm{(3nd axiom)}\\
&\to W \indep M|A,U,X \quad \quad \textrm{(Assumption \ref{consist})}\\
&\to W \indep (M,A)|U,X \quad \quad (W\indep A|U,X \textrm{ and 4th axiom)}\\
&\to W \indep A|M,U,X \quad \quad \textrm{(3rd axiom)}
\end{align*}

\subsection*{Proof of Theorem \ref{id_or}}

\begin{proof}
Following the proof of Theorem 1 in \cite{miao2018identifying}:
\begin{align*}
E(Y|z,A=1,m,x)&=\int E(Y|u,z,A=1,m,x)dF(u|z,A=1,m,x) \quad \quad (\textrm{Tower rule})\\
&=\int E(Y|u,A=1,m,x)dF(u|z,A=1,m,x) \quad \quad  (Z\indep Y|M,A,U,X).%\\
%&=\int h_1(w,A=1,M,X)dF(w|Z,A=1,M,X)\\
\end{align*}
Also. %the equality (\ref{fred1}) and Assumption \ref{p_proxy} implies that 
\begin{align*}
&E(Y|z,A=1,m,x)\\
&=\int h_1(w,m,x)dF(w|z,A=1,m,x) \quad \quad (\textrm{by (\ref{fred1})})\\
&=\int \int h_1(w,m,x)dF(w|u,z,A=1,m,x)dF(u|z,A=1,m,x)\quad \quad (\textrm{Tower rule})\\
&=\int \int h_1(w,m,x)dF(w|u,A=1,m,x)dF(u|z,A=1,m,x) \quad \quad (Z\indep W|M,A,U,X)
\end{align*}
and by the completeness condition \ref{complete}.1, result (\ref{int1}) follows.

Then
\begin{align*}
E\{h_1(W,M,x)|z,A=0,x\}&= \int E\{h_1(W,M,x)|u,z,A=0,x\}dF(u|z,A=0,x) \quad \quad (\textrm{Tower rule})\\
&=\int E\{h_1(W,M,x)|u,A=0,x\}dF(u|z,A=0,x) \quad \quad (Z\indep (W,M)|A,U,X) %
\end{align*}
and 
\begin{align*}
E\{h_1(W,M,x)|z,A=0,x\}&=\int h_0(w,x)dF(w|z,A=0,x) \quad \quad (\textrm{by (\ref{fred2})}) \\
&=\int \int h_0(w,x)dF(w|z,u,A=0,x)dF(u|z,A=0,x) \quad \quad (\textrm{Tower rule})\\
&=\int \int h_0(w,x)dF(w|u,A=0,x)dF(u|z,A=0,x) \quad \quad (Z \indep W|A,U,X)
%&=\int h_1(w,A=1,M,X)dF(w|Z,A=1,M,X)\\
\end{align*}
and by the completeness condition \ref{complete}.2, we have  
\begin{align}\label{int2}
E\{h_1(W,M,X)|U,A=0,X\}=\int h_0(w,X)dF(w|U,A=0,X).
\end{align}
 %Note that by the graphoid axioms, Assumption \ref{p_proxy} implies the $W\indep Z|A,U,X$.

Finally, it follows from the above results that 
\begin{align*}
\psi&= \int \int E[Y\{1,M(0)\}|u,x]dF(u|x)dF(x)\\
&= \int \int \int  E\{Y(1,m)|u,M(0)=m,x]dF_{M(0)}(m|u,x)dF(u|x)dF(x)\quad \quad(\textrm{Tower rule})\\
&= \int \int \int  E\{Y(1,m)|u,x]dF_{M(0)}(m|u,x)dF(u|x)dF(x) \quad \quad (\textrm{Assumption \ref{cwa}}) \\
&= \int \int \int  E\{Y(1,m)|u,A=1,x]dF_{M(0)}(m|u,x)dF(u|x)dF(x) \quad \quad (\textrm{Assumption \ref{lce}.1})\\
&= \int \int \int  E\{Y(1,m)|u,A=1,m,x]dF_{M(0)}(m|u,x)dF(u|x)dF(x) \quad \quad (\textrm{Assumption \ref{lce}.2})\\
&= \int \int \int  E\{Y|u,A=1,m,x]dF_{M(0)}(m|u,x)dF(u|x)dF(x) \quad \quad (\textrm{Assumption \ref{consist}})\\
&= \int \int \int  E\{Y|u,A=1,m,x]dF_{M(0)}(m|u,A=0,x)dF(u|x)dF(x) \quad \quad (\textrm{Assumption \ref{lce}.3})\\
&=\int \int \int E(Y|u,A=1,m,x)dF(m|u,A=0,x)dF(u|x)dF(x)\quad \quad (\textrm{Assumption \ref{consist}})\\
&=\int \int \int \int h_1(w,m,x)dF(w|u,A=1,m,x)dF(m|u,A=0,x)dF(u|x)dF(x) \quad \quad (\textrm{by (\ref{int1})})\\
&=\int \int \int \int h_1(w,m,x)dF(w,m|u,A=0,x)dF(u|x)dF(x) \quad \quad (W\indep A|M,U,X)\\
&=\int \int \int  h_0(w,x)dF(w|u,A=0,x)dF(u|x)dF(x) \quad \quad (\textrm{by (\ref{int2})})\\
&=\int \int \int h_0(w,x)dF(w|u,x)dF(u|x)dF(x) \quad \quad (W\indep A|U,X)\\
&=\int \int h_0(w,x)dF(w|x)dF(x) \quad \quad(\textrm{Tower rule})
\end{align*}
%where the third equality follows by $W\indep A|U,M,X$ the fifth follows by $W\indep A|U,X$.
\end{proof}
where we index by $M(0)$ to indicate that the integration is done over the counterfactual $M(0)$ rather than the observed $M$.

\subsection*{Proof of Theorem \ref{id_alt}}

\begin{assumption}\label{complete2}(Completeness)
\begin{enumerate}
\item For any function $g(u)$, if $E\{g(U)|W=w,A=1,M=m,X=x\}=0$ for any $w$, $m$ and $x$ almost surely, then $g(U)=0$ almost surely.
\item For any function $g(u)$, if $E\{g(U)|W=w,A=0,X=x\}=0$ for any $w$ and $x$ almost surely, then $g(U)=0$ almost surely. %(CHECK)
%\item For any function $g(w)$,  $E\{g(W)|Z=z,A=1,M=m,X=x\}=0$ for any $z$, $m$ and $x$ almost surely, then $g(W)=0$ almost surely.
%\item For any function $g(w)$,  $E\{g(W)|Z=z,A=0,X=x\}=0$ for any $z$ and $x$ almost surely, then $g(W)=0$ almost surely. %(CHECK)
\end{enumerate}
\end{assumption}

\begin{proof}
Result (\ref{cui_res}) follows directly from Theorem 2.2 of \cite{cui2020semiparametric}, but is reproduced here for completeness. 
First, 
\begin{align*}
 \frac{1}{f(A=0|w,x)}&=\int \frac{1}{f(A=0|w,x)} dF(u|w,x)\quad \\
 &=\int \frac{1}{f(A=0|u,w,x)} dF(u|w,A=0,x)\quad \\
 &=\int \frac{1}{f(A=0|u,x)} dF(u|w,A=0,x) \quad \quad (W\indep A|U,X)
\end{align*}
and also 
\begin{align*}
 \frac{1}{f(A=0|w,x)}&=E\{q_0(Z,x)|w,A=0,x\} \quad \quad (\textrm{by (\ref{fred3})})\\
 &=\int \int q_0(Z,x)dF(z|u,w,A=0,x) dF(u|w,A=0,x) \quad \quad(\textrm{Tower rule})\\
  &=\int \int q_0(Z,x)dF(z|u,A=0,x) dF(u|w,A=0,x) \quad \quad (Z\indep W|A,U,X)
\end{align*}
and by Assumption \ref{complete2}.2, we have (\ref{cui_res}). Further
%By Assumptions \ref{consist}-\ref{p_proxy} and the results of Theorem \ref{id_or}, 
\begin{align*}
\psi&=\int \int \int E(Y|u,A=1,m,x)dF(m|u,A=0,x)dF(u|x)dF(x)\quad \quad (\textrm{Assumptions \ref{consist}, \ref{lce} and \ref{cwa}})\\
%&=\int \int \int E(Y|u,A=1,m,x)dF(m|u,A=0,x)dF(u|x)dF(x)\quad \quad (\textrm{Assumptions \ref{consist}, \ref{lce} and \ref{cwa}})\\
&=\int \int \int E(Y|u,A=1,m,x)dF(m|u,A=0,x)\frac{f(A=0|u,x)}{f(A=0|u,x)}dF(u|x)dF(x)\\
&=E\left\{\frac{I(A=0)}{f(A=0|U, X)}E(Y|U,A=1,M,X)\right\}\quad \quad(\textrm{Tower rule})\\
&=E\left[I(A=0)E\{q_0(Z,X)|U,A=0,X\}E(Y|U,A=1,M,X)\right] \quad \quad (\textrm{by (\ref{cui_res})})\\
&=E\left[I(A=0)E\{q_0(Z,X)|U,A=0,M,X\}E(Y|U,A=1,M,X)\right] \quad \quad (Z\indep M|A,U,X)\\
&=E\left[I(A=0)E\{q_0(Z,X)|U,A,M,X\}E(Y|U,A=1,M,X)\right] \\
&=E\left\{I(A=0)q_0(Z,X)E(Y|U,A=1,M,X)\right\}\quad \quad(\textrm{Tower rule})\\
&=E\left[I(A=0)q_0(Z,X)E\{h_1(W,M,X)|U,A=1,M,X\}\right]\quad \quad (\textrm{by (\ref{int1})})\\
&=E\left[I(A=0)q_0(Z,X)E\{h_1(W,M,X)|U,A=0,M,X\}\right] \quad \quad (W\indep A|M,U,X)\\
&=E\left[I(A=0)q_0(Z,X)E\{h_1(W,M,X)|Z,U,A=0,M,X\}\right] \quad \quad (Z\indep W|M,A,U,X)\\
&=E\left[I(A=0)q_0(Z,X)h_1(W,M,X)\right] \quad \quad(\textrm{Law of iterated expectation.})
%\psi&=\int \int \int \frac{I(a=0)}{f(a|u,x)}E(Y|u,1,m,x)dF(u,z,a,m|x)dF(x)\\
%&=\int \int \int E(Y|u,A=1,m,x)dF(m|u,A=0,x)dF(u,x)\\
\end{align*}
%since $M\indep Z|A,U,X$, $W \indep A|M,U,X$ and $W\indep Z|A,M,U,X$.

For the second part of Theorem \ref{id_alt}, %by Bayes rule and Assumption \ref{p_proxy},
\begin{align*}
&E\{q_0(Z,x)|w,A=0,m,x\}\frac{f(A=0|w,m,x)}{f(A=1|w,m,x)}\\
&=\frac{1}{f(A=1|w,m,x)} \int E\{q_0(Z,x)|u,w,A=0,m,x\}f(A=0|w,m,x)dF(u|w,A=0,m,x) \\&(\textrm{Tower rule})\\
& = \frac{1}{f(A=1|w,m,x)} \int E\{q_0(Z,x)|u,w,A=0,m,x\}f(A=0|u,w,m,x)dF(u|w,m,x) \\
& = \int E\{q_0(Z,x)|u,w,A=0,m,x\}\frac{f(A=0|u,w,m,x)}{f(A=1|u,w,m,x)}dF(u|w,A=1,m,x) \\
& = \int E\{q_0(Z,x)|u,w,A=0,m,x\}\frac{f(A=0|u,m,x)}{f(A=1|u,m,x)}dF(u|w,A=1,m,x)\quad \quad (W\indep A|M,U,X)\\
& = \int E\{q_0(Z,x)|u,A=0,m,x\}\frac{f(A=0|u,m,x)}{f(A=1|u,m,x)}dF(u|w,A=1,m,x)\quad \quad (Z\indep W|M,A,U,X)\\
\end{align*}
where  for the third equality, we use that 
\[f(U|W,A=1,M,X)f(A=1|W,M,X)=f(A=1|U,W,M,X)f(U|W,M,X)\]
and so 
\[\frac{f(U|W,A=1,M,X)}{f(A=1|U,W,M,X)}=\frac{f(U|W,M,X)}{f(A=1|W,M,X)}.\]
Also 
%equality (\ref{fred4}) implies that 
\begin{align*}
&E\{q_0(Z,x)|w,A=0,m,x\}\frac{f(A=0|w,m,x)}{f(A=1|w,m,x)}\\
%&\int E\{q_0(Z,x)|u,A=0,m,x\}\frac{f(A=0|u,m,x)}{f(A=1|u,m,x)}dF(u|w,A=1,m,x)\\
&= \int q_1(z,m,x)dF(z|w,A=1,m,x) \quad \quad (\textrm{by (\ref{fred4})})\\
&= \int \int q_1(z,m,x)dF(z|u,w,A=1,m,x)dF(u|w,A=1,m,x)\quad \quad(\textrm{Tower rule})\\
&= \int \int q_1(z,m,x)dF(z|u,A=1,m,x)dF(u|w,A=1,m,x)\quad \quad (Z\indep W|M,A,U,X)
\end{align*}
Hence 
\begin{align}\label{int4}
&E\{q_0(Z,X)|U,A=0,M,X\}\frac{f(A=0|U,M,X)}{f(A=1|U,M,X)}= E\left\{q_1(Z,M,X)|U,A=1,M,X\right\} 
\end{align}
follows from Assumption \ref{complete2}.1. Finally,
%Also, if we invoke Assumptions \ref{consist}-\ref{p_proxy}, then 
\begin{align*}
\psi&=E\left\{\frac{I(A=0)}{f(A=0|U, X)}E(Y|U,A=1,M,X)\right\}\quad \quad (\textrm{Assumptions \ref{consist}, \ref{lce} and \ref{cwa}})\\
&=E\left[I(A=0)E\{q_0(Z,X)|U,A=0,X\}E(Y|U,A=1,M,X)\right]\quad \quad (\textrm{by }(\ref{cui_res}))\\
&=E\left[I(A=0)E\{q_0(Z,X)|U,A=0,M,X\}E(Y|U,A=1,M,X)\right]\quad \quad (Z\indep M|A,U,X)\\
&=E\left\{I(A=0)q_0(Z,X)E(Y|U,A=1,M,X)\right\} \quad \quad (\textrm{Tower rule})\\
&=E\left\{I(A=0)q_0(Z,X)E(Y|Z,U,A=1,M,X)\right\} \quad \quad  (Z\indep Y|M,A,U,X)\\
%&=\int \int \int q_0(z,x)E(Y|u,1,m,x)f(0|u,z,x)dF(m|u,0,z,x)dF(u,z|x)dF(x)\\
&=\int \int q_0(z,x)E(Y|z,u,A=1,m,x)f(A=0|u,m,z,x)dF(u,m,z|x)dF(x) \quad \quad(\textrm{Tower rule}) \\
&=\int \int q_0(z,x)E(Y|u,A=1,m,x)f(A=0|u,m,z,x)dF(u,m,z|x)dF(x)  \quad \quad (Y\indep Z|M,A,U,X)\\
%&=\int \int E\{q_0(z,x)|u,A=0,m,x\}E(Y|u,A=1,m,x)f(A=0|u,m,x)dF(u,m|x)dF(x) \\&\quad \quad(\textrm{Tower rule}) \\
&=E\left[E\{q_0(Z,X)|U,A=0,M,X\}E(Y|U,A=1,M,X)f(A=0|U,M,X)\right]\quad \quad(\textrm{Tower rule})\\
&=E\left[I(A=1)Y E\{q_0(Z,X)|U,A=0,M,X\}\frac{f(A=0|U,M,X)}{f(A=1|U,M,X)}\right]\quad \quad(\textrm{Tower rule})\\
&=E\left[I(A=1)Y E\{q_1(Z,M,X)|U,A=1,M,X\}\right]\quad \quad (\textrm{by (\ref{int4})})\\
&=E\left[I(A=1)Y q_1(Z,M,X)\right] \quad(\textrm{Tower rule}) .
\end{align*}
%since $M\indep Z | A,U,X$ and $Y \indep Z|A,M,U$.
\end{proof}

\subsection*{Proof of Theorem \ref{eif}}

\begin{proof}

In order to first obtain an influence function for $\psi$ under the semiparametric model $\mathcal{M}_{sp}$, one can find the mean zero random variable $G$ for which
\begin{align*}
\frac{\partial \psi_t}{\partial t}|_{t=0}=E\{GS(O;t)\}|_{t=0}
\end{align*}
where $S(O;t)=\partial \log f(O;t)/\partial t$. Here, we take a derivative with respect to the parameter $t$ indexing a one-dimensional parametric submodel in $\mathcal{M}_{sp}$, which returns true density $f(O)$ at $t=0$.

After noting the moment restrictions
\begin{align*}%\label{moment}
E\{Y-h_1(W,M,X)|Z,A=1,M,X\}&=0\\
E\{h_1(W,M,X)-h_0(W,X)|Z,A=0,X\}&=0
\end{align*}
implied by (\ref{fred1}) and (\ref{fred2}), then letting $\mathcal{E}_1=Y-h_1(W,M,X)$ and $\mathcal{E}_0=h_1(W,M,X)-h_0(W,X)$, it follows that 
\begin{align}
&E\left\{\frac{\partial}{\partial t} h_{0_t}(W,X)|_{t=0}\bigg|Z,A=0,X\right\}\nonumber\\
&=E\left\{\mathcal{E}_0S(W,M|Z,A=0,X)|Z,A=0,X\right\}+E\left\{\frac{\partial}{\partial t} h_{1_t}(W,M,X)|_{t=0}\bigg|Z,A=0,X\right\}\label{srec1}
\end{align}
and
\begin{align}
&E\left\{\frac{\partial}{\partial t} h_{1_t}(W,M,X)|_{t=0}\bigg|Z,A=1,M,X\right\}=E\left\{\mathcal{E}_1S(Y,W|Z,A=1,M,X)|Z,A=1,M,X\right\}.\label{srec2}
\end{align}
%For the above expressions to be well-defined, note that the bridge functions $h_1$ and $h_0$ are required to exist 

Then
\begin{align*}
\frac{\partial \psi_t}{\partial t}|_{t=0}&=\frac{\partial}{\partial t}E_t[h_0(W,X)]|_{t=0}=E[h_0(W,X)S(W,X)]+E\left\{\frac{\partial}{\partial t} h_{0_t}(W,X)|_{t=0}\right\}\\
&=E[\{h_0(W,X)-\psi\}S(O)]+E\left\{\frac{\partial}{\partial t} h_{0_t}(W,X)|_{t=0}\right\}
\end{align*}
For the second term on the right hand side of the final equality,
\begin{align*}
&E\left\{\frac{\partial}{\partial t} h_{0_t}(W,X)|_{t=0}\right\}
=E\left\{ \frac{I(A=0)}{f(A=0|W,X)}  \frac{\partial}{\partial t} h_{0_t}(W,X)|_{t=0}\right\}\\
=&E\left\{ I(A=0)E\{q_0(Z,X)|W,A=0,X\}\frac{\partial}{\partial t} h_{0_t}(W,X)|_{t=0}\right\}\\
=&E\left\{ I(A=0)q_0(Z,X) \frac{\partial}{\partial t} h_{0_t}(W,X)|_{t=0}\right\}\\
=&E\left[ I(A=0)q_0(Z,X) E\left\{\frac{\partial}{\partial t} h_{0_t}(W,X)|_{t=0}\bigg|Z,A=0,X\right\}\right]\\
=&E\left[ I(A=0)q_0(Z,X) E\left\{\mathcal{E}_0S(W,M|Z,A=0,X)|Z,A=0,X\right\}\right]\\
&\quad+E\left[ I(A=0)q_0(Z,X) E\left\{\frac{\partial}{\partial t} h_{1_t}(W,M,X)|_{t=0}\bigg|Z,A=0,X\right\}\right].\end{align*}
and 
\begin{align*}
&E\left[ I(A=0)q_0(Z,X) E\left\{ \frac{\partial}{\partial t} h_{1_t}(W,M,X)|_{t=0}\bigg|Z,A=0,X\right\}\right]\\
%&=\int \int f(0|z,x)q_0(z,x) \left\{ \int  \frac{\partial}{\partial t} h_{1_t}(w,m,x)|_{t=0}dF(m,w|z,0,x)\right\}dF(z|x)d(x)\\
%&=\int \int \int  f(0|w,z,x)q_0(z,x)\frac{\partial}{\partial t} h_{1_t}(w,m,x)|_{t=0}dF(m|w,z,0,x)dF(w,z|x)dF(x)\\
%&=\int \int f(0|w,z,m,x)q_0(z,x)\frac{\partial}{\partial t} h_{1_t}(w,m,x)|_{t=0}dF(w,z,m|x)dF(x)\\
&=E\left\{I(A=0)q_0(Z,X)\frac{\partial}{\partial t} h_{1_t}(W,M,X)|_{t=0}\right\}\\
%&=\int f(A=0|z,w)q_0(z) \frac{\partial}{\partial t} h_{1_t}(w,m)|_{t=0}\frac{f(A=0|m,z,w)}{f(A=0|z,w)}dP(m,z,w)\\
%&=E\left\{q_0(Z) f(A=0|M,Z,W) \frac{\partial}{\partial t} h_{1_t}(W,M)|_{t=0}\right\}\\
&=E\left[f(A=0|W,M,X)E\{q_0(Z,X)|W,A=0,M,X\}\frac{\partial}{\partial t} h_{1_t}(W,M,X)|_{t=0}\right]\\
&=E\left[I(A=1)\frac{f(A=0|W,M,X)}{f(A=1|W,M,X)}E\{q_0(Z,X)|A=0,M,W\}\frac{\partial}{\partial t} h_{1_t}(W,M,X)|_{t=0}\right]\\
&=E\left[I(A=1)E\{q_1(Z,M,X)|W,A=1,M,X\} \frac{\partial}{\partial t}h_{1_t}(W,M,X)|_{t=0}\right]\\
&=E\left\{I(A=1)q_1(Z,M,X) \frac{\partial}{\partial t}h_{1_t}(W,M,X)|_{t=0}\right\}\\
&=E\left[I(A=1)q_1(Z,M,X) E\left\{ \frac{\partial}{\partial t}h_{1_t}(W,M,X)|_{t=0}\bigg|Z,A=1,M,X\right\}\right]\\
&=E\left[I(A=1)q_1(Z,M,X) E\left\{\mathcal{E}_1S(Y,W|Z,A=1,M,X)|Z,A=1,M,X\right\}\right]
\end{align*}

Putting this together, we have 
\begin{align*}
&E\left\{\frac{\partial}{\partial t} h_{0_t}(W,X)|_{t=0}\right\}\\
&=E\left\{I(A=0)q_0(Z,X)\mathcal{E}_0S(W,M|Z,A=0,X)\right\}\\
&\quad +E\left\{I(A=1)q_1(Z,M,X) \mathcal{E}_1S(Y,W|Z,A=1,M,X)\right\}\\
&=E\left[ I(A=0)q_0(Z,X)\mathcal{E}_0E\{S(O)|W,M,Z,A,X\}\right]\\
&\quad -E\left[ I(A=0)q_0(Z,X)\mathcal{E}_0E\{S(O)|Z,A,X\}\right]\\
&\quad +E\left[ I(A=1)q_1(Z,M,X) \mathcal{E}_1E\{S(O)|Y,W,Z,A,M,X\}\right]\\
&\quad -E\left[ I(A=1)q_1(Z,M,X) \mathcal{E}_1E\{S(O)|Z,A,M,X\}\right]\\
&=E\left[ I(A=0)q_0(Z,X)\mathcal{E}_0S(O)\right]+E\left[ I(A=1)q_1(Z,M,X) \mathcal{E}_1S(O)\right]
\end{align*}
To show that
\begin{align*}
&I(A=1)q_1(Z,M,X)\mathcal{E}_1+I(A=0)q_0(Z,X)\mathcal{E}_0+h_0(W,X)-\psi
\end{align*}
 is the efficient influence function, we need to show that it belongs to the tangent space implied by the restrictions (\ref{srec1}) and (\ref{srec2}) on the scores. Specifically, by (\ref{srec1}) and (\ref{srec2}) the tangent space is comprised of the set of scores $S(O)\in L_2(O)$ that satisfy
 \begin{align*}
 &E\left\{\mathcal{E}_1S(Y,W|Z,A=1,M,X)|Z,A=1,M,X\right\} \in R(T_1)\\
 &E\left\{\mathcal{E}_0S(W,M|Z,A=0,X)|Z,A=0,X\right\} \in R(T_0)
 \end{align*}
 where $R()$ denotes the range space of an operator. Following the proof of Theorem 3 in \citet{ying2021proximal}, by Assumption \ref{bound} holding at the true law, the tangent space equals $L_2(O)$ and the main result follows.

\end{proof}

\subsection*{Proof of Theorem \ref{res_if_q1}}

\begin{proof}
From (\ref{fred4}), we have that 
\[E\left[q_1(Z,M,X;\gamma_1)-\kappa(W,M,X;\gamma_0)\bigg|W,A=1,M,X\right]=0\]
where
\[\kappa(W,M,X;\gamma_0)=E\{q_0(Z,X;\gamma_0)|W,A=0,M,X\}\frac{f(A=0|W,M,X)}{f(A=1|W,M,X)}\]
and therefore 
\begin{align*}
\partial E_t\bigg[& \left\{q_1(Z,M,X;\gamma_{1_t})-\kappa_t(W,M,X;\gamma_{0_t})\right\}d_1(W,M,X)\bigg|A=1\bigg]/\partial t|_{t=0}=0
\end{align*}
for any $d_1(W,M,X)$ of the same dimension as $\gamma_1$, where
\[\kappa_t(W,M,X;\gamma_{0_t})=E_t\{q_0(Z,X;\gamma_{0_t})|W,A=0,M,X\}\frac{f_t(A=0|W,M,X)}{f_t(A=1|W,M,X)}.\]

After some algebra and re-arrangement of terms, 
\begin{align*}
&-\frac{1}{f(A=1)}E \left\{\frac{\partial q_1(Z,M,X;\gamma_1)}{\partial \gamma_1}Ad_1(W,M,X) \right\}\frac{\partial \gamma_{1_t}}{\partial t}|_{t=0}\\
&= \frac{1}{f(A=1)} E\left[A \left\{q_1(Z,M,X;\gamma_1)-\kappa(W,M,X;\gamma_0)\right\}d_1(W,M,X)S(O)\right]\\
&\quad - \frac{1}{f(A=1)} E\left[\frac{\partial q_0(Z,X;\gamma_0)}{\partial \gamma_0}(1-A)d_1(W,M,X)\right]\frac{\partial \gamma_{0_t}}{\partial t}|_{t=0}  \\
&\quad - \frac{1}{f(A=1)}E\left[(1-A)q_0(Z,X;\gamma_{0})S(Z|W,A,M,X)d_1(W,M,X)\right]\\
&\quad -E\left[\frac{ \partial f_t(A=0|W,M,X)/\partial t \vert_{t=0}}{f(A=1|W,M,X)}E\{q_0(Z,X;\gamma_{0})|W,A=0,M,X\}d_1(W,M,X)\bigg|A=1\right]\\
&\quad +E\left[\frac{ \partial f_t(A=1|W,M,X)/\partial t \vert_{t=0}}{f^2(A=1|W,M,X)}f(A=0|W,M,X)E\{q_0(Z;
\gamma_{0})|W,A=0,M,X\}d_1(W,M,X)\bigg|A=1\right]
\end{align*}
For the third term on the right hand side of the equality,
\begin{align*}
&-\frac{1}{f(A=1)}E\left[(1-A)q_0(Z,X;\gamma_{0})S(Z|W,A,M,X)d_1(W,M,X)\right]\\
&=-\frac{1}{f(A=1)}E\left((1-A)\left[q_0(Z,X;\gamma_{0})-E\{q_0(Z,X;\gamma_{0})|W,A=0,M,X\}\right]d_1(W,M,X)S(O)\right)
\end{align*}
For the penultimate term, we have
\begin{align*}
&-E\left[\frac{ \partial f_t(A=0|W,M,X)/\partial t_{t=0}}{f(A=1|W,M,X)}E\{q_0(Z,X;\gamma_{0})|W,A=0,M,X\}d_1(W,M,X)\bigg|A=1\right]\\
&=-E\left[\frac{E\{(1-A)S(A|W,M,X)|W,M,X\}}{f(A=1|W,M,X)}E\{q_0(Z,X;\gamma_{0})|W,A=0,M,X\}d_1(W,M,X)\bigg|A=1\right]\\
&=-\frac{1}{f(A=1)}E\left[E\{(1-A)S(A|W,M,X)|W,M,X\}E\{q_0(Z,X;\gamma_{0})|W,A=0,M,X\}d_1(W,M,X)\right]\\
&=-\frac{1}{f(A=1)}E\left[(1-A)S(A|W,M,X)q_0(Z,X;\gamma_{0})d_1(W,M,X)\right]\\
&=-\frac{1}{f(A=1)}E\left[\left\{(1-A)-f(A=0|W,M,X)\right\}E\{q_0(Z,X;\gamma_{0})|W,A=0,M,X\}d_1(W,M,X)S(O)\right].
\end{align*}
For the final term
\begin{align*}
&E\left[\frac{ \partial f_t(A=1|W,M,X)/\partial t_{t=0}}{f^2(A=1|W,M,X)}f(A=0|W,M,X)E\{q_0(Z,X;\gamma_{0})|W,A=0,M,X\}d_1(W,M,X)\bigg|A=1\right]\\
%&=E\left[\frac{E\{AS(A|W,M,X)|W,M,X\}}{f^2(A=1|W,M,X)}f(A=0|W,M,X)E\{q_0(Z,X;\gamma_{0})|W,A=0,M,X\}d_1(W,M,X)\bigg|A=1\right]\\
%&=\frac{1}{f(A=1)}E\left[\frac{E\{AS(A|W,M,X)|W\}}{f(A=1|W,M,X)}f(A=0|W,M,X)E\{q_0(Z,X;\gamma_{0})|W,A=0,M,X\}d_1(W,M,X)\right]\\
&=\frac{1}{f(A=1)}E\left[AS(A|W,M,X)\frac{f(A=0|W,M,X)}{f(A=1|W,M,X)}E\{q_0(Z,X;\gamma_{0})|W,A=0,M,X\}d_1(W,M,X)\right]\\
&=\frac{1}{f(A=1)}E\bigg[\left\{\frac{A-f(A=1|W,M,X)}{f^2(A=1|W,M,X)}\right\}\kappa(W,M,X;\gamma_0)d_1(W,M,X)S(O)\bigg].
\end{align*}

Putting this all together, it follows that 
\begin{align*}
&-\frac{1}{f(A=1)}E \left\{\frac{\partial q_1(Z,M,X;\gamma_1)}{\partial \gamma_1}Ad_1(W,M,X) \right\}\frac{\partial \gamma_{1_t}}{\partial t}|_{t=0}\\
&= \frac{1}{f(A=1)} E\left[A \left\{q_1(Z,M,X;\gamma_1)-\kappa(W,M,X;\gamma_0)\right\}d_1(W,M,X)S(O)\right]\\
&\quad - \frac{1}{f(A=1)} E\left[\frac{\partial q_0(Z,X;\gamma_0)}{\partial \gamma_0}(1-A)d_1(W,M,X)\right]\frac{\partial \gamma_{0_t}}{\partial t}|_{t=0}  \\
&\quad -\frac{1}{f(A=1)}E\left((1-A)\left[q_0(Z;
\gamma_{0})-E\{q_0(Z,X;\gamma_{0})|W,A=0,M,X\}\right]d_1(W,M,X)S(O)\right)\\
&\quad -\frac{1}{f(A=1)}E\left[\left\{(1-A)-f(A=0|W,M,X)\right\}E\{q_0(Z,X;\gamma_{0})|W,A=0,M,X\}d_1(W,M,X)S(O)\right]\\
&\quad +\frac{1}{f(A=1)}E\bigg[\left\{\frac{A-f(A=1|W,M,X)}{f^2(A=1|W,M,X)}\right\}\kappa(W,M,X;\gamma_0)d_1(W,M,X)S(O)\bigg]\\
&=\frac{1}{f(A=1)} E\left[\left\{Aq_1(Z,M,X;\gamma_1)-(1-A)q_0(Z,X;\gamma_{0})\right\}d_1(W,M,X)S(O)
\right]\\
&\quad - \frac{1}{f(A=1)} E\left[\frac{\partial q_0(Z,X;\gamma_0)}{\partial \gamma_0}(1-A)d_1(W,M,X)\right]\frac{\partial \gamma_{0_t}}{\partial t}|_{t=0}.
\end{align*}
%where the main result follows by the application of Theorem 3.2 in \citet{cui2020semiparametric}.
\end{proof}

\subsection*{Proof of Theorem \ref{res_mr}}

\begin{proof}
Let $\beta^*_1$, $\beta^*_0$, $\gamma^*_1$ and $\gamma^*_0$ refer to the population limits of the estimators $\hat{\beta}_1$,  $\hat{\beta}_0$, $\hat{\gamma}_1$ and $\hat{\gamma}_0$. Then if $h_1(W,M,X;\beta^*_1)$ and $h_0(W,X;\beta^*_0)$ are correctly specified,
\begin{align*}
&E\left[I(A=1)q_1(Z,M,X;\gamma^*_1)\{Y-h_1(W,M,X;\beta^*_1)\}\right]\\&\quad+E[I(A=0)q_0(Z,X;\gamma^*_0)\{h_1(W,M,X;\beta^*_1)-h_0(W,X;\beta^*_0)\}]+E\{h_0(W,X;\beta^*_0)\}-\psi\\
&=E\left[I(A=1)q_1(Z,M,X;\gamma^*_1)\left\{E(Y|Z,A=1,M,X)-\int h_1(w,M,X;\beta^*_1)dF(w|Z,A=1,M,X)\right\}\right]\\
&\quad+E\left(I(A=0)q_0(Z,X;\gamma^*_0)\left[E\{h_1(W,M,X;\beta^*_1)|Z,A=0,X\}-\int h_0(w,X;\beta^*_0) dF(w|Z,A=0,X)\right]\right)\\&\quad+E\{h_0(W,X;\beta^*_0)\}-\psi\\
&=0
\end{align*}
by virtue of (\ref{fred1}) and (\ref{fred2}). If $h_1(W,M,X;\beta_1^*)$ and $q_0(Z,X;\gamma^*_0)$ are instead correctly specified,
\begin{align*}
&E\left[I(A=1)q_1(Z,M,X;\gamma^*_1)\{Y-h_1(W,M,X;\beta^*_1)\}\right]\\&\quad+E[I(A=0)q_0(Z,X;\gamma^*_0)\{h_1(W,M,X;\beta^*_1)-h_0(W,X;\beta^*_0)\}]+E\{h_0(W,X;\beta^*_0)\}-\psi\\
&=E[I(A=0)q_0(Z,X;\gamma^*_0)h_1(W,M,X;\beta^*_1)]-E[I(A=0)q_0(Z,X;\gamma^*_0)h_0(W,X;\beta^*_0)]\\&\quad+E\{h_0(W,X;\beta^*_0)\}-\psi
\end{align*}
Now by (\ref{fred2}) and (\ref{fred3}),
\begin{align*}
&E[I(A=0)q_0(Z,X;\gamma^*_0)h_1(W,M,X;\beta^*_1)]\\
&=E[I(A=0)q_0(Z,X;\gamma^*_0)E\{h_1(W,M,X;\beta^*_1)|Z,A=0,X\}]\\
&=E[I(A=0)q_0(Z,X;\gamma^*_0)E\{h_0(W,X)|Z,A=0,X\}]\\
&=E[I(A=0)E\{q_0(Z,X;\gamma^*_0)|W,A=0,X\}h_0(W,X)]\\
&=E\left\{\frac{I(A=0)}{f(A=0|W,X)}h_0(W,X)\right\}\\
&=\psi
\end{align*}
and 
\begin{align*}
&E\{h_0(W,X;\beta^*_0)\}-E[I(A=0)q_0(Z,X;\gamma^*_0)h_0(W,X;\beta^*_0)]\\
%&=E\{h_0(W,X;\beta^*_0)\}-E[I(A=0)E\{q_0(Z,X;\gamma^*_0)|Z,A=0,X\}h_0(W,X;\beta^*_0)]\\
&=E\{h_0(W,X;\beta^*_0)\}-E\left\{\frac{I(A=0)}{f(A=0|W,X)}h_0(W,X;\beta^*_0)\right\}\\
&=E\{h_0(W,X;\beta^*_0)\}-E\left\{h_0(W,X;\beta^*_0)\right\}\\
&=0
\end{align*}
Finally, if $q_1(Z,M,X;\gamma^*_1)$ and $q_0(Z,X;\gamma^*_0)$ are instead correctly specified,
\begin{align*}
&E\left[I(A=1)q_1(Z,M,X;\gamma^*_1)\{Y-h_1(W,M,X;\beta^*_1)\}\right]\\&\quad+E[I(A=0)q_0(Z,X;\gamma^*_0)\{h_1(W,M,X;\beta^*_1)-h_0(W,X;\beta^*_0)\}]+E\{h_0(W,X;\beta^*_0)\}-\psi\\
&=E\{I(A=1)q_1(Z,M,X;\gamma^*_1)Y\}-E\{I(A=1)q_1(Z,M,X;\gamma^*_1)h_1(W,M,X;\beta^*_1)\}\\&\quad+E[I(A=0)q_0(Z,X;\gamma^*_0)h_1(W,M,X;\beta^*_1)]-\psi.
\end{align*}
By (\ref{fred1}), (\ref{fred3}) and (\ref{fred4}),
\begin{align*}
&E\{I(A=1)q_1(Z,M,X;\gamma^*_1)Y\}\\
&=E\{I(A=1)q_1(Z,M,X;\gamma^*_1)E(Y|Z,A=1,M,X)\}\\
&=E\{I(A=1)q_1(Z,M,X;\gamma^*_1)h_1(W,M,X)\}\\
&=E[I(A=1)E\{q_1(Z,M,X;\gamma^*_1)|W,A=1,M,X\}h_1(W,M,X)]\\
&=E\left[I(A=1)E\{q_0(Z,X;\gamma^*_0)|W,A=0,M,X\}\frac{f(A=0|W,M,X)}{f(A=1|W,M,X)}h_1(W,M,X)\right]\\
&=E\left[E\{q_0(Z,X;\gamma^*_0)|W,A=0,M,X\}f(A=0|W,M,X)h_1(W,M,X)\right]\\
&=E\left[I(A=0)q_0(Z,X;\gamma^*_0)h_1(W,M,X)\right]\\
&=\psi
\end{align*}
and also 
\begin{align*}
&E[I(A=0)q_0(Z,X;\gamma^*_0)h_1(W,M,X;\beta^*_1)]-E\{I(A=1)q_1(Z,M,X;\gamma^*_1)h_1(W,M,X;\beta^*_1)\}\\
&=E[I(A=0)q_0(Z,X;\gamma^*_0)h_1(W,M,X;\beta^*_1)]\\&\quad-E[I(A=1)E\{q_1(Z,M,X;\gamma^*_1)|Z,A=1,M,X\}h_1(W,M,X;\beta^*_1)]\\
&=E[I(A=0)q_0(Z,X;\gamma^*_0)h_1(W,M,X;\beta^*_1)]\\&\quad-E\left[I(A=1)E\{q_0(Z,X;\gamma^*_0)|W,A=0,M,X\}\frac{f(A=0|W,M,X)}{f(A=1|W,M,X)}h_1(W,M,X;\beta^*_1)\right]\\
&=E[I(A=0)q_0(Z,X;\gamma^*_0)h_1(W,M,X;\beta^*_1)]-E\left[I(A=0)q_0(Z,X;\gamma^*_0)h_1(W,M,X;\beta^*_1)\right]\\
&=0
\end{align*}
and the first result in Theorem \ref{res_mr} follows.

%The efficient influence function under the semiparametric model $\mathcal{M}_{union}$ coincides with the efficient influence function under $\mathcal{M}_{sp}$, following a result in \citet{robins_comments_2001}. 

To show (local) semiparametric efficiency, we will show that 
$\hat{\psi}_{P-MR}$ has an influence function equal to  $IF_\psi$ under $\mathcal{M}_1\cap \mathcal{M}_2 \cap \mathcal{M}_3$. Following a Taylor expansion and standard $M$-estimation arguments, we have that 
\begin{align*}
&\sqrt{n}(\hat{\psi}_{P-MR}-\psi)\\&=\frac{1}{\sqrt{n}}\sum^n_{i=1}A_iq_1(Z_i,M_i,X_i;\gamma^*_1)\{Y_i-h_1(W_i,M_i,X_i;\beta^*_1)\}\\&+\frac{1}{\sqrt{n}}\sum^n_{i=1}(1-A_i)q_0(Z_i,X_i;\gamma^*_0)\{h_1(W_i,M_i,X_i;\beta^*_1)-h_0(W_i,X_i;\gamma^*_0)\}\\&+\frac{1}{\sqrt{n}}\sum^n_{i=1}h_0(W_i,X_i;\gamma^*_0)-\psi\\
&+E\left[ I(A=1)\frac{\partial q_1(Z,M,X;\gamma^*_1)}{\partial \gamma^*_1}\{Y-h_1(W,M,X;\beta^*_1)\}\right]\sqrt{n}(\hat{\gamma}^*_1-\gamma^*_1)\\
&+E\left[ I(A=0)\frac{\partial q_0(Z,X;\gamma^*_0)}{\partial \gamma^*_0}\{h_1(W,M,X;\beta^*_1)-h_0(W,X;\beta^*_0)\} \right]\sqrt{n}(\hat{\gamma}^*_0-\gamma^*_0)\\
&+E\left[\frac{\partial h_1(W,M,X;\beta^*_1)}{\partial \beta^*_1} \{I(A=0)q_0(Z,X;\gamma^*_0)- I(A=1)q_1(Z,M,X;\gamma^*_1)\}\right]\sqrt{n}(\hat{\beta}^*_1-\beta^*_1)\\
&+E\left[\frac{\partial h_0(W,X;\beta^*_0)}{\partial \beta^*_0} \{1-I(A=0)q_0(Z,X;\gamma^*_0)\}\right]\sqrt{n}(\hat{\beta}^*_0-\beta^*_0)\\
&+o_p(1)
\end{align*}
Following the previous arguments in this section, under model $\mathcal{M}_{union}$,
\begin{align*}
E\left[ I(A=1)\frac{\partial q_1(Z,M,X;\gamma^*_1)}{\partial \gamma^*_1}\{Y-h_1(W,M,X;\beta^*_1)\}\right]&=0\\
E\left[ I(A=0)\frac{\partial q_0(Z,X;\gamma^*_0)}{\partial \gamma^*_0}\{h_1(W,M,X;\beta^*_1)-h_0(W,X;\beta^*_0)\} \right]&=0\\
E\left[\frac{\partial h_1(W,M,X;\beta^*_1)}{\partial \beta^*_1} \{I(A=0)q_0(Z,X;\gamma^*_0)- I(A=1)q_1(Z,M,X;\gamma^*_1)\}\right]&=0\\
E\left[\frac{\partial h_0(W,X;\beta^*_0)}{\partial \beta^*_0} \{1-I(A=0)q_0(Z,X;\gamma^*_0)\}\right]&=0
\end{align*}
which completes the proof. 
\end{proof}

\subsection*{Randomised trials}\label{rct}

\begin{figure}
\centering
	\begin{tikzpicture}[scale=1]
	\node[] (a) at (0, 0)   {$A$};
	\node[] (m) at (4, 1.5)   {$M$};
	\node[] (y) at (8, 0)   {$Y$};
	\node[draw,circle] (u) at (8,4) 	 {$U$};
	\node[] (w) at (10,2.5) 	 {$W$};
	\node[] (z) at (4,2.5) 	 {$Z$};
	\node[] (x) at (7,2.5) 	 {$X$};

	\path[->] (a) edge node {} (y);
	\path[->] (a) edge node {} (m);
	%\path[->] (a) edge node {} (z);
	\path[->] (m) edge node {} (y);
	\path[->] (x) edge node {} (m);
	\path[->] (x) edge node {} (y);
	\path[->] (x) edge node {} (z);
	\path[->] (x) edge node {} (w);
	%\path[->] (u) edge node {} (m);
	\path[->] (u) edge node {} (z);
	\path[->] (u) edge node {} (x);
	\path[->] (u) edge node {} (m);
	%\path[->] (z) edge node {} (m);
	\path[->] (u) edge node {} (w);
	%\path[->] (w) edge node {} (y);
	\path[->] (u) edge node {} (y);
	\end{tikzpicture}
\caption{Causal diagram with unmeasured mediator-outcome confounding.}
\label{proxy_rct}
\end{figure}
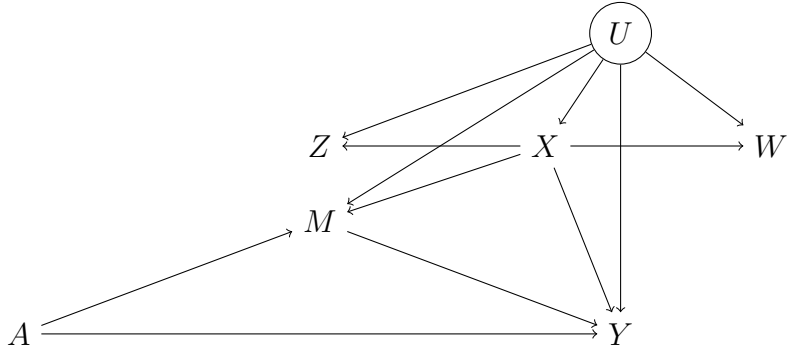

    We will consider briefly estimation of $\psi$ in a setting where the exposure $A$ is randomised at baseline. In that case, by virtue of randomisation, there can be no confounding of either the exposure-outcome or exposure-mediator relationship. Nevertheless, we cannot discount the possibility of unmeasured mediator-outcome confounders. Figure \ref{proxy_rct} describes a setting where there is an unmeasured common cause of $M$ and $Y$, in addition to a measured common cause $X$, but where again we will make progress thanks to proxy variables $Z$ and $W$. Although in Figure \ref{proxy_rct} encodes the assumption that the are no unmeasured common causes of $A$ and $Y$, or $A$ and $M$, to be sufficiently general the following results allow for $X$ to influence $A$.

%In what follows, we can exploit the fact that the randomisation probability $f(A=0)=\pi$ is known, in order to construct a doubly robust estimator of $\psi$.

\iffalse
\begin{assumption}\label{bound_rct}(Additional regularity conditions)
\begin{enumerate}
%\item Let $\Psi_1(z,m,x)\equiv E\left[\{Y-h_1(W,m,x)\}^2|Z=z,M=m,X=x\right]$ and \\$\Psi_0(z,x)\equiv E\left[\{h_1(W,m,x)-h_0(W,x)\}^2|Z=z,X=x\right]$. Then we assume that 
%$0< \inf_{z,m,x}\Psi_1(z,m,x)\leq \sup_{z,m,x}\Psi_1(z,m,x)<\infty$ and $0< \inf_{z,x}\Psi_0(z,x)\leq \sup_{z,x}\Psi_0(z,x)<\infty$.
%\item Let $\Psi(z,m,x)\equiv E\left[\{Y-h_1(W,m,x)\}^2|Z=z,A=1,M=m,X=x\right]$. Then we assume that 
%$0< \inf_{z,m,x}\Psi(z,m,x)\leq \sup_{z,m,x}\Psi(z,m,x)<\infty$.
%\item Let $\Gamma_0(w,x)\equiv E\left[\left\{q_0(Z,x)-\frac{1}{f(A=0|w,x)}\right\}^2|W=w,X=x\right]$ and \\$\Gamma_1(w,m,x)\equiv E\left[\left\{q_1(Z,m,x)-\frac{f(A=0|w,m,x)}{f(A=1|w,m,x)}\right\}^2|W=w,M=m,X=x\right]$. Then we assume that $0< \inf_{w,x}\Gamma_0(w,x)\leq \sup_{w,x}\Gamma_0(w,x)<\infty$ and $0< \inf_{w,m,x}\Gamma_1(w,m,x)\leq \sup_{w,m,x}\Gamma_1(w,m,x)<\infty$.
\item Let $T :L_2(W,A,M,X)\to L_2(Z,M,X)$ denote the operator given by \\$T(g)\equiv E\{g(W,M,X)|Z,A=1,M,X\}$ and the adjoint $T' :L_2(Z,M,X)\to L_2(W,M,X)$ denote $T'(g)\equiv E\{g(Z,M,X)|W,A=1,M,X\}$. $D(T)$ denotes the domain space of $T$. Then we require that%, $A^{\perp}$ is the orthocomplement of $A$ and $cl(A)$ is the closure of $A$. 
\begin{align*}
&T'\Psi \left(I(A=0)q_0(Z,X)E\left[\{Y-h_1(W,M,X)\}\{h_1(W,M,X)-\eta_0(W,X)\}|Z,A=1,M,X\right]\right)\\&\in D(T'\psi T)^{-1} \quad \textrm{and} \quad
T'\{I(A=1)q_1(Z,M,X)\} \in D(T'\psi T)^{-1}
\end{align*}
\end{enumerate}
\end{assumption}
\fi

\begin{theorem}\label{theo_rct}
(Part a)  We will assume that there exist proxies that satisfy Assumptions \ref{p_proxy}.1, \ref{p_proxy}.3, \ref{p_proxy}.4 and $U\indep A|W,X$, and that there exist $h_1$, and $q_1$ that satisfy (\ref{fred1}) and 
\begin{align}\label{fred5}
%E(Y|Z,A=1,M,X)&=\int h_1(w,M,X)dF(w|Z,A=1,M,X)\\
%E\{h_1(W,M,X)|Z_0,A=0,X\}&=\int h_0(w_0,X)dF(w_0|Z_0,A=0,X)\\
%\frac{1}{f(A=0|W_0,X)}&=E\{q_0(Z_0,X)|W_0,A=0,X\}\\
\frac{f(A=0|W,M,X)}{f(A=1|W,M,X)}&=E\left\{q_1(Z,M,X)|W,A=1,X\right\}. 
\end{align}
Suppose furthermore that $Y(a,m)\indep A|X$ for $a=0,1$ and each $m\in \mathcal{S}$, $M(a)\indep A|X$ for $a=0,1$, $Pr(A=a|X)>0$ almost surely for $a=0,1$ and that 
Assumptions \ref{consist}, \ref{pos}.1, \ref{lce}.2, \ref{cwa},\ref{complete}.1 and \ref{complete2}.1 hold. %, where \ref{pos}.2, \ref{lce}.1 and \ref{lce}.3 now hold conditional on $Z$ and $X$, but not $U$. 
If we define
\[\eta_0(W,X)=\int h_1(W,m,X)dF(m|W,A=0,M,X)\]
then, we again have results (\ref{int1}),
\begin{align}\label{int_rct}
%E(Y|Z,A=1,M,X)&=\int h_1(w,M,X)dF(w|Z,A=1,M,X)\\
%E\{h_1(W,M,X)|Z_0,A=0,X\}&=\int h_0(w_0,X)dF(w_0|Z_0,A=0,X)\\
%\frac{1}{f(A=0|W_0,X)}&=E\{q_0(Z_0,X)|W_0,A=0,X\}\\
\frac{f(A=0|U,M,X)}{f(A=1|U,M,X)}&=E\left\{q_1(Z,M,X)|U,A=1,M,X\right\} 
\end{align}
and we have the three following representations of the proximal mediation formula 
\begin{align*}
\psi&=\int \int  \eta_0(w,x)dF(w|x)dF(x)\\
&=\int  \int \frac{I(a=0)}{f(a=0|x)}h_1(w,m,x)dF(w,z,a,m|x)dF(x)\\
&=\int  \int \frac{I(a=1)}{f(a=1|x)}q_1(z,m,x)ydF(y,z,a,m|x)dF(x)
\end{align*}
(Part b) Consider a semiparametric model $\mathcal{M}^*_{sp}$ where $h_1$ is assumed to exist at every data law under the model but the observed data distribution is otherwise left unrestricted. Furthermore, $q_1$ is assumed to exist at the true data-generating law and Assumption \ref{complete}.1 is assumed to hold.
%Under the conditions from Part a, $h_1$ and $q_1$ that solve (\ref{fred1}), and (\ref{fred5}) are uniquely identified. 
Then a valid influence function for $\psi$ under $\mathcal{M}^*_{sp}$ is equal to 
\begin{align*}
IF^*_{\psi}=&\frac{I(A=1)}{f(A=0|X)}q_1(Z,M,X)\{Y-h_1(W,M,X)\}\\&+\frac{I(A=0)}{f(A=0|X)}\{h_1(W,M,X)-\eta_0(W,X)\}+\eta_0(W,X)-\psi.
\end{align*}
Furthermore, in the submodel where $h_1$ and $q_1$ are unique and Assumption \ref{bound}.1 holds, the efficiency bound for $\psi$ is $E(IF_{\psi}^{*2})$. The efficiency bound is also unchanged in the case that $f(A=0|X)$ is known.
\end{theorem}

%\subsection*{Proof of Theorem \ref{theo_rct}}

\begin{proof}
For Part a), results (\ref{int1}) follows from the proof of Theorem \ref{id_or}, by Assumptions \ref{consist}, \ref{lce}.2, \ref{p_proxy}.1, \ref{p_proxy}.3, \ref{p_proxy}.4, \ref{complete}.1 and the existance of a bridge function that satisfies (\ref{fred1}). Similarly, (\ref{int_rct}) follows from the proof of Theorem \ref{id_alt}, by Assumptions \ref{consist}, \ref{lce}.2, \ref{p_proxy}.1, \ref{p_proxy}.3, \ref{p_proxy}.4, \ref{complete}.1 and the existence of a bridge function that satisfies (\ref{fred5}).

Then
\begin{align*}%\label{tmp1a}
E(Y|u,A=1,m,x)=\int h_1(w,M,X)dF(w|u,A=1,M,X)
\end{align*}
following the proof of Theorem \ref{id_or},
and 
\begin{align*}
\psi&=\int \int \int E(Y|u,A=1,m,x)dF(m|u,A=0,x)dF(u|x)dF(x) \quad \quad \textrm{(Assumptions \ref{consist}-\ref{cwa})}\\
&=\int \int \int \int h_1(w,m,x)dF(w|u,A=1,m,x)dF(m|u,A=0,x)dF(u|x)dF(x) \quad \quad \textrm{(by (\ref{int1}))} \\
&=\int \int \int \int h_1(w,m,x)dF(w|u,A=0,m,x)dF(m|u,A=0,x)dF(u|x)dF(x) \quad \quad (W\indep A|M,U,X) \\
&=\int \int \int \int h_1(w,m,x)dF(m|u,A=0,m,x)dF(w|u,A=0,x)dF(u|x)dF(x)\\
&=\int \int \int \int h_1(w,m,x)dF(m|u,A=0,m,x)dF(w|u,x)dF(u|x)dF(x) \quad \quad  W\indep A |U,X\\
&=\int \int \int \int h_1(w,m,x)dF(m|u,A=0,m,x)dF(u|w,x)dF(w|x)dF(x)\\
&=\int \int \int \int h_1(w,m,x)dF(m|u,A=0,w,x)dF(u|w,A=0,x)dF(w|x)dF(x)\quad \quad \textrm{($U\indep A|W,X$)}  \\
&=\int \int \int  h_1(w,m,x)dF(m|A=0,w,x)dF(w|x)dF(x)
%&=\int \int \int h_1(w,m,x)dF(m,w|u,0,x)dF(u|0,x)dF(x)\\
%&=\int \int  h_1(w,m,x)dF(w,m|0,x)dF(x).
%&=\int \int  h_1(w,m,x)dF(m|w,0,x)dF(w|x)dF(x)
\end{align*}
and we note that  $W\indep A|U,X$ and $W\indep A|M,U,X$ follows from Assumptions \ref{consist}, \ref{p_proxy}.3 and \ref{p_proxy}.4, by application of the graphoid axioms.

Part b) follows along the lines of similar arguments to the proof of Theorem \ref{eif} and is omitted for brevity. We nevertheless clarify that $IF^*_\psi$ is orthogonal to the scores $S(A|X)$. First, using the law of iterated expectations,
\begin{align*}
&E\left[\frac{A}{f(A=1|X)}q_1(Z,M,X)\{Y-h_1(W,M,X)\}S(A|X)\right]\\
&=E\left[\frac{A}{f(A=1|X)}q_1(Z,M,X)E\{Y-h_1(W,M,X)|Z,A=1,M,X\}S(A|X)\right]=0
\end{align*}
by virtue of (\ref{fred1}). Next, 
\begin{align*}
&E\left[\frac{(1-A)}{f(A=0|X)}\{h_1(W,M,X)-\eta_0(W,X)\}S(A|X)\right]\\
&=E\left[\frac{(1-A)}{f(A=0|X)}E\{h_1(W,M,X)-\eta_0(W,X)|A=0,W,X\}S(A|X)\right]=0
\end{align*}
by the definition of $\eta_0(W,X)$. Finally, by the (conditional) randomisation of $A$,
\begin{align*}
&E\left[\{\eta_0(W,X)-\psi\}S(A|X)\right]\\
&E\left[\{\eta_0(W,X)-\psi\}E\{S(A|X)|X\}\right]=0
\end{align*}
and hence we have shown that $E\{IF^*_\psi S(A|X)\}=0$.

%where the third equality follows by $W\indep A|U,M,X$ and $A\indep U|X$.

%\frac{(1-A)}{f(A=0|X)}\{h_1(W,M,X)-\eta_0(X)\}+\eta_0(X)-\psi

\end{proof}

\subsection*{Model compatibility}\label{mod_comp}

By Bayes' theorem 
\begin{align*}
\frac{f(M|U,A=0,X)}{f(M|U,A=1,X)}=\frac{f(A=0|U,M,X)}{f(A=1|U,M,X)}\times \frac{f(A=1|U,X)}{f(A=0|U,X)}. 
\end{align*}
Suppose that \[M|U,A,X\sim \mathcal{N}(\tau_0+\tau_aA+\tau_u U+\tau_x X,\sigma^2_{m|u,a,x});\] then 
one can show that 
\[\log\frac{f(M|U,A=0,X)}{f(M|U,A=1,X)}=\frac{\tau_a}{\sigma^2_{m|u,a,x}}\left\{
\frac{\tau_a}{2}-(M-\tau_0-\tau_u U-\tau_x X)\right\} \] which is linear in $U$ and $X$. Further, if 
\[\frac{1}{f(A=0|U,X)}=1+\exp\{-(\alpha_{0,0}+\alpha_{0,u}U+\alpha_{0,x}X)\}\] then 
\begin{align*}
\log \frac{f(A=0|U,M,X)}{f(A=1|U,M,X)}&=\log \frac{f(M|U,A=0,X)}{f(M|U,A=1,X)} + \log \frac{f(A=0|U,X)}{f(A=1|U,X)}\\
&=\frac{\tau_a}{\sigma^2_{m|u,a,x}}\left\{
\frac{\tau_a}{2}-(M-\tau_0-\tau_u U-\tau_x X)\right\}\\&\quad + \alpha_{0,0}+\alpha_{0,u}U+\alpha_{0,x}X\\
&= \alpha_{1,0}+\alpha_{1,m}M+\alpha_{1,u}U+\alpha_{1,x}X
\end{align*}
where 
\[\alpha_{1,0}=\frac{\tau_a}{\sigma^2_{m|u,a,x}}\left(
\frac{\tau_a}{2}+\tau_0\right)+ \alpha_{0,0},\]
$\alpha_{1,m}=-\tau_a/\sigma^2_{m|u,a,x}$, $\alpha_{1,u}=\tau_a\tau_u/\sigma^2_{m|u,a,x}+\alpha_{0,u}$ and $\alpha_{1,x}=\tau_a\tau_x/\sigma^2_{m|u,a,x}+\alpha_{0,x}$.

Next, suppose that  $Z|U,A,X\sim \mathcal{N}(\epsilon_{0}+\epsilon_{u} U+\epsilon_{a}A+\epsilon_{x} X,\sigma^2_{z|u,a,x})$, and consider the choice of bridge function
\[q_0(Z,X)=1+\exp\{-(\gamma_{0,0}+\gamma_{0,z}Z+\gamma_{0,x}X)\},\]
then
\begin{align*}
E\{q_0(Z,X)|U,A=0,X\}&= 1+\exp(-\gamma_{0,0}-\gamma_{0,x}X)\int \exp(-\gamma_{0,z}z) dF(z|U,A=0,X)\\
&=1+\exp\left\{-\gamma_{0,0}-\gamma_{0,x}X-\gamma_{0,z}(\epsilon_{0}+\epsilon_{u} U+\epsilon_{x} X)+\frac{\gamma^2_{0,z}\sigma^2_{z|u,a,x}}{2}\right\}
\end{align*}
%such that equality (\ref{fred3}) holds,  
such that for 
\begin{align*}
\gamma_{0,0}&=\alpha_{0,0}-\frac{\alpha_{0,u}}{\epsilon_{u}}\left\{\epsilon_{0}-\frac{(\alpha_{0,u}/\epsilon_{u})\sigma^2_{z|u,a,x}}{2}\right\}\\
\gamma_{0,z}&=\frac{\alpha_{0,u}}{\epsilon_{u}}\\
\gamma_{0,x}&=\alpha_{0,x}-\frac{\alpha_{0,u}\epsilon_{x}}{\epsilon_{u}}
\end{align*}
we have the equality
\[\frac{1}{f(A=0|U,X)}=\int q_0(z,x)dF(z|U,A=0,X)\]
To show this choice of $q_0(Z,X)$ also lead to equality (\ref{fred3}), since $W\indep A|U,X$ we have that 
\begin{align*}
\frac{1}{f(A=0|W,X)}&=\int \frac{1}{f(A=0|u,X)}dF(u|W,A=0,X)\\
&=1+\exp(-\alpha_{0,0}-\alpha_{0,x}X)\int \exp (-\alpha_{0,u}u)dF(u|W,A=0,X)
%&=1+\exp\left\{\alpha_{0,0}+\alpha_{0,x}X+\alpha_{0,u}E(U|W,A=0,X)+\frac{\alpha^2_{0,u}\sigma^2_{U|0,X}}{2}\right\}
\end{align*}
Also, 
\begin{align*}
&E\{q_0(Z,X)|W,A=0,X\}=E[E\{q_0(Z,X)|U,A=0,X\}|W,A=0,X]\\
&=1+\exp(-\alpha_{0,0}-\alpha_{0,x}X)\int \exp (-\alpha_{0,u}u)dF(u|W,A=0,X)
\end{align*}
by $Z\indep W|U,A,X$ (which follows from Assumption (\ref{p_proxy})) and equality (\ref{fred3}).

Moving onto $q_1$, consider the choice of bridge function
\begin{align*}
q_1(Z,M,X)=&\exp\{\gamma_{1,0}+\gamma_{1,z}Z+\gamma_{1,m}M+\gamma_{1,x}X\}\\
&\times \left[1+\exp(-\gamma_{0,0}-\gamma_{0,z}Z-\gamma_{0,x}X)\exp\left\{\gamma_{0,z}\left(\epsilon_a+\gamma_{1,z}\sigma^2_{z|u,a,x}\right)\right\}\right]
\end{align*}
Noting that by $Z\indep M|U,A,X$, 
\begin{align*}
&E\{q_1(Z,M,X)|U,A=1,M,X\}\\&=\exp\left\{\gamma_{1,0}+\frac{\gamma^2_{1,z}\sigma^2_{z|u,a,x}}{2}+\gamma_{1,z}E(Z|U,A=1,X)+\gamma_{1,m}M+\gamma_{1,x}X\right\}\\
&\quad+\exp\left\{(\gamma_{1,0}-\gamma_{0,0})+\gamma_{0,z}\left(\epsilon_a+\gamma_{1,z}\sigma^2_{z|u,a,x}\right)+\gamma_{1,m}M+(\gamma_{1,x}-\gamma_{0,x})X\right\}\\
&\quad\quad \times \int \exp\{(\gamma_{1,z}-\gamma_{0,z})z\}dF(z|U,A=1,X)\\
&=\exp\left\{\gamma_{1,0}+\frac{\gamma^2_{1,z}\sigma^2_{z|u,a,x}}{2}+\gamma_{1,z}E(Z|U,A=1,X)+\gamma_{1,m}M+\gamma_{1,x}X\right\}\\
&\quad+\exp\left\{(\gamma_{1,0}-\gamma_{0,0})+\gamma_{0,z}\left(\epsilon_a+\gamma_{1,z}\sigma^2_{z|u,a,x}\right)+\gamma_{1,m}M+(\gamma_{1,x}-\gamma_{0,x})X\right\}\\
&\quad\quad \times \exp\left\{(\gamma_{1,z}-\gamma_{0,z})E(Z|U,A=1,X)+\frac{\gamma^2_{1,z}\sigma^2_{z|u,a,x}}{2}+\frac{\gamma^2_{0,z}\sigma^2_{z|u,a,x}}{2}-\gamma_{1,z}\gamma_{0,z}\sigma^2_{z|u,a,x}\right\}\\
&=\exp\left\{\gamma_{1,0}+\frac{\gamma^2_{1,z}\sigma^2_{z|u,a,x}}{2}+\gamma_{1,z}E(Z|U,A=1,X)+\gamma_{1,m}M+\gamma_{1,x}X\right\}\\
&\quad \times \left[1+\exp\left\{-\gamma_{0,0}-\gamma_{0,x}X-\gamma_{0,z}(\epsilon_{0}+\epsilon_{0} U+\epsilon_{x} X)+\frac{\gamma^2_{0,z}\sigma^2_{z|u,a,x}}{2}\right\}\right]\\
&=\exp\left\{\gamma_{1,0}+\frac{\gamma^2_{1,z}\sigma^2_{z|u,a,x}}{2}+\gamma_{1,z}E(Z|U,A=1,X)+\gamma_{1,m}M+\gamma_{1,x}X\right\}E\{q_0(Z,X)|U,A=0,X\}
\end{align*}
Therefore, it follows that this choice of $q_1(Z,M,X)$ satisfies the equality 
\begin{align*}
&E\{q_0(Z,X)|U,A=0,M,X\}\frac{f(A=0|U,M,X)}{f(A=1|W,M,X)}= E\left\{q_1(Z,M,X)|U,A=1,M,X\right\} 
\end{align*}
at the parameter values 
\begin{align*}
\gamma_{1,0}&=\alpha_{1,0}-\frac{\alpha_{1,u}}{\epsilon_{u}}\left\{\frac{\alpha_{1,u}\sigma^2_{z|u,a,x}}{2\epsilon_{u}}+\epsilon_0+\epsilon_a\right\}\\
\gamma_{1,m}&=\alpha_{1,m}\\
\gamma_{1,z}&=\frac{\alpha_{1,u}}{\epsilon_{u}}\\
\gamma_{1,x}&=\alpha_{1,x}-\frac{\alpha_{1,u}\epsilon_{x}}{\epsilon_{u}}
\end{align*}
It furthermore follows that under the constraint that 
\[\epsilon_a+\frac{\sigma^2_{z|u,a,x}}{\epsilon_u} \left\{\frac{\tau_a\tau_u}{\sigma^2_{m|u,a,x}}+\alpha_{0,u}\right\}=0\]
that this choice of bridge function simplifies to 
\[q_0(Z,X)\exp\{\gamma_{1,0}+\gamma_{1,z}Z+\gamma_{1,m}M+\gamma_{1,x}X\}.\]
It also follows from previous reasoning that this choice of bridge also leads to the equality (\ref{fred4}).

\newpage

\subsection*{Additional simulation results}

\begin{table}[htbp]
\centering
\caption{Simulation results from experiments 5-9. Exp: experiment; Est: estimator; MSE: mean squared error; Bias: Monte Carlo bias; Med. Bias: Median estimate minus true value;  Coverage: 95\% confidence interval (CI) coverage; Mean Length: average 95\% CI length; Med. length: median 95\% CI length.}
%\begin{adjustbox}{width=\columnwidth,center}
\resizebox{\textwidth}{!}{%
\begin{tabular}{llllllll}
  \hline
Exp & Est & Bias & Med. Bias & MSE & Coverage & Mean Length & Med. Length \\ 
  \hline
5 & $\hat{\theta}_{OLS}$ & 0.00 & 0.00 & 0.01 & 0.96 & 0.38 & 0.38 \\ 
   & $\hat{\theta}_{P-IPW}$ & 0.00 & 0.00 & 0.01 & 0.96 & 0.48 & 0.48 \\ 
   & $\hat{\theta}_{P-hybrid}$ & 0.00 & 0.00 & 0.01 & 0.96 & 0.48 & 0.48 \\ 
   & $\hat{\theta}_{P-OR}$ & 0.00 & 0.00 & 0.01 & 0.96 & 0.48 & 0.48 \\ 
   & $\hat{\theta}_{P-MR}$ & 0.00 & 0.00 & 0.01 & 0.96 & 0.48 & 0.48 \\ 
  6 & $\hat{\theta}_{OLS}$ & 0.31 & 0.31 & 0.11 & 0.12 & 0.39 & 0.39 \\ 
   & $\hat{\theta}_{P-IPW}$ & 0.40 & 0.40 & 0.17 & 0.03 & 0.54 & 0.41 \\ 
   & $\hat{\theta}_{P-hybrid}$ & 0.40 & 0.40 & 0.17 & 0.03 & 0.53 & 0.41 \\ 
   & $\hat{\theta}_{P-OR}$ & 0.40 & 0.40 & 0.17 & 0.04 & 0.49 & 0.41 \\ 
   & $\hat{\theta}_{P-MR}$ & 0.40 & 0.40 & 0.17 & 0.03 & 0.54 & 0.41 \\ 
  7 & $\hat{\theta}_{OLS}$ & 0.26 & 0.27 & 0.08 & 0.24 & 0.39 & 0.39 \\ 
   & $\hat{\theta}_{P-IPW}$ & -0.34 & -0.34 & 0.14 & 0.37 & 43.97 & 0.57 \\ 
   & $\hat{\theta}_{P-hybrid}$ & -0.34 & -0.34 & 0.14 & 0.33 & 9.83 & 0.54 \\ 
   & $\hat{\theta}_{P-OR}$ & -0.34 & -0.33 & 0.13 & 0.33 & 7.11 & 0.54 \\ 
   & $\hat{\theta}_{P-MR}$ & -0.34 & -0.34 & 0.14 & 0.37 & 43.97 & 0.57 \\ 
  8 & $\hat{\theta}_{OLS}$ & 0.36 & 0.36 & 0.14 & 0.05 & 0.39 & 0.39 \\ 
   & $\hat{\theta}_{P-IPW}$ & 0.07 & 0.15 & 1.51 & 0.98 & $>$1000 & 16.08 \\ 
   & $\hat{\theta}_{P-hybrid}$ & -0.03 & 0.11 & 42.53 & 0.99 & $>$1000 & 6.74 \\ 
   & $\hat{\theta}_{P-OR}$ & -0.16 & 0.08 & 59.27 & 1.00 & $>$1000 & 9.65 \\ 
   & $\hat{\theta}_{P-MR}$ & -0.11 & 0.11 & 51.33 & 0.98 & $>$1000 & 18.24 \\ 
  9 & $\hat{\theta}_{OLS}$ & 0.14 & 0.14 & 0.04 & 0.87 & 0.6 & 0.60 \\ 
   & $\hat{\theta}_{P-IPW}$ & 0.04 & 0.05 & 0.60 & 0.97 & $>$1000 & 14.57 \\ 
   & $\hat{\theta}_{P-hybrid}$ & 0.28 & 0.04 & 84.41 & 0.99 & $>$1000 & 1.87 \\ 
   & $\hat{\theta}_{P-OR}$ & 0.28 & 0.04 & 76.73 & 0.99 & $>$1000 & 2.38 \\ 
   & $\hat{\theta}_{P-MR}$ & 0.05 & 0.05 & 1.52 & 0.98 & $>$1000 & 16.20 \\ 
   \hline
\end{tabular}
}
%\end{adjustbox}
\label{results_tab2}
\end{table}

\section*{Data analysis: additional information}

The information below follows the AGReMA statement \citep{lee2021guideline}, which are guidelines for 
good practice for conducting and reporting mediation analysis in randomised trials and observational studies.

\begin{enumerate}
    \item \textit{Title}: The Job Corps study.
    \item \textit{Abstract}: To what extend is the effect of attending class at the beginning of the study on criminal activity mediated by employment?
    \item \textit{Background and rationale}: Using data from the Job Corps study, \citet{schochet2001national} and \citet{schochet2008does} consider the total effect of program assignment on multiple outcomes; their results suggest that being assigned to the program leads to a reduction in criminal activity for some years after the program. \citet{flores2009identification} and \citet{huber2014identifying} see positive direct effects of assignment/attendence on earnings/health outcomes with the mediator employment. Using an IV design, \citet{frolich2017direct} find evidence of an indirect rather than direct effect of attendance on earnings, as mediated by employment. Less is known about the role of employment as a mediator in the relationship between program attendance and criminal activity.
    \item \textit{Objectives}: Estimate the direct and indirect effects of program attendance (rather than assignment) on number of arrests at year four, with the number of hours worked at year two being the mediator.
    \item \textit{Study registration}: No protocol or study registration available for the mediation analysis.
    \item \textit{Study design}: Job Corps was a randomised controlled trial carried out at 119 separate centres, across 48 states and the District of Columbia. Eligible individuals between the ages of 16-24 were assigned at random either to receive an offer of participation in the Job Corps program or to be declined access to participation in the program for the following three years. Survey data on relevant variables was collected from participants around 2 and 4 years after randomisation. See \citet{schochet2001national} and \citet{schochet2008does} for further details.
    \item \textit{Participants}: Data on 15,386 individuals was collected at baseline; however, many individuals failed to enroll in the program and dropped out of the study. As in other closely related analyses \citep{colangelo2020double,huber2020direct,singh2021kernel}, we restricted our analysis to the 10,775 individuals for which the mediator and outcome were observed.
    \item \textit{Sample size}: No sample size calculation was conducted for the mediation analysis.
    \item \textit{Effects of interest}: Total effects and natural direct and indirect effects.
    \item \textit{Assumed causal model}: See Figure A.1(a).
    \item \textit{Causal assumptions}: See Assumptions \ref{consist}-\ref{complete} in the main manuscript and Assumption \ref{complete2}. 
    \item \textit{Measurement and measurement levels}: Exposure variable $A$: was any time spent in academic or vocational classes in the 12 months following randomisation according to the survey (binary)? Mediator: proportion of weeks employed in the second year, according to the survey (numerican variable, in percentages). Outcome: number of separate arrests in the fourth year after randomisation, according to the survey (numeric). Confounders: gender (binary), age (numeric), ethnicity (categorical), education status (categorical), native English (binary), marital status (categorical), has children (binary), ever worked (binary), average weekly earnings (numeric), head of household (binary), designated for nonresidential slot (binary), total household gross income (categorical), dad did not work when 14 (binary), welfare receipt during childhood (categorical), poor/fair general health status (binary), received AFDC each month (binary), received public assistance each month (binary), received food stamples (binary), physical emotional problems (binary), ever taken illegal drugs that are not marijuana or hallucinogens (binary), extent of smoking (categorical), extent of alcohol consumption (categorical), ever arrested (binary), times in prison (categorical). 
    
    %g2: ever arrested
    %g7: received good stanmpsh2
    %h2: physical/emotional problems
    %h29: ever taken illegal drugs (not mj or psych)
    %h5: extent of smoking
    %h7: extend of alcohol consumption. 
    %i1: ever arrested
    %i10: times in prison

    %l=cbind(female, age, race_white, race_black, race_hispanic,   educ_geddiploma, educ_hsdiploma, ntv_engl, marstat_divorced, marstat_separated, marstat_livetogunm, marstat_married, haschldY0, everwkd,  mwearn, hohhd0, nonres,g10, g10missdum, work_dad_didnotwork, g2, g5, g7, welfare_child, welfare_childmissdum, h1_fair_poor, h2, h29, h5, h5missdum, h7, h7missdum, i1, i10)[e12missdum==0,] #update

   % \item \textit{Measurement levels}: Exposure: binary; 0 (zero hours of class attended attended), 1 (>0 hours of class attended). Mediator: numeric variable, in percentages. Outcome:  numeric variable.
    \item \textit{Statistical methods}: See main manuscript for outline of analysis and details on models/estimators chosen. We removed several variables that were subject to large amounts of missingness (extent of marijuana use, extend of hallucinogen use), since we hypothesised that any confounding would be accounted for by other variables detailing substance use. We also excluded years of education, household size, mum's years of education and dad's years of education, as these also contained missing values and were strongly correlated with covariates with complete data. As in previous analyses of this sample \citep{colangelo2020double,huber2020direct,singh2021kernel}, the missing indicator method was used for the remaining covariates that had missing values; see \citet{groenwold2012missing} for a discussion of the limitations of this approach. We also removed any individuals with incomplete data on the proxies; a complete case analysis is always unbiased when data is missing completely at random, but can be biased under certain missing at random and missing not at random mechanisms. However, since the proportion of individuals removed was small (less than 3\% of the total sample size), we hypothesised that any resulting bias (and efficiency loss) would be negligible. All analyses were performed using R. 
    \item \textit{Sensitivity analysis}: A sensitivity analysis for the causal identification assumptions was beyond the scope of the article. \citet{huber2020direct} did not find strong evidence of the dependence of missing values of outcome and mediator on treatment. 
    \item \textit{Ethical approval}: See \citet{schochet2001national}.
    \item \textit{Participants}: See \citet{schochet2008does}, \citet{huber2020direct}, \citet{huber2020direct} and other previous analyses of the Job Corps study.
    \item \textit{Outcomes and estimates}: See main manuscript, and table \ref{apptab}.
    \item \textit{Sensitivity parameters}:  No sensitivity analysis performed. 
    \item \textit{Limitations}: Similar to analyses in \citet{colangelo2020double,huber2020direct,singh2021kernel}, results may be sensitive to post-treatment confounding, given that the mediator was assessed two years after randomisation. Results may also be sensitive to bias due to loss to follow up. 
    \item \textit{Interpretation}: See main manuscript.
    \item \textit{Implications}: There was no strong evidence that participation in the Job studies program affected the number of arrests; either directly or via an effect on employment. However, based on results e.g. in \citet{huber2014identifying}, its possible that a dose-response relationship occurred for the direct effect, such that in those who participated, attending more hours of class led to a reduction in the number of arrests (outside of any mediation via employment). 
    \item \textit{Funding and role of sponsor}: Study sponsors are discussed e.g. in \citet{schochet2001national}. See Acknowledgements in main paper for information on grants that supported the authors of the mediation analysis. Grant funders had no role in the conduct of the study, writing of the manuscript, and decision to submit for publication.
    \item \textit{Conflicts of interest and financial disclosures}: None to report. 
    \item \textit{Data and code}: Data is freely available at http://qed.econ.queensu.ca/jae/datasets/hsu001/. Code is available upon request.
    %Code for implementing estimators is available as part of Supplementary Material.
\end{enumerate}

\begin{table}[ht]
\caption{Additional results from the analysis of the Job Corps study. CI: confidence interval.}
\centering
\begin{adjustbox}{width=\textwidth}
\begin{tabular}{rrrrrrrrrrrrrrrr}
  \hline
 &  P-IPW & 95\% CI  & P-Hybrid & 95\% CI  & P-OR & 95\% CI   \\ 
  \hline
$E[Y\{1,M(0)\}]-[Y\{0,M(0)\}]$ & 0.0946 & 0.0251,0.1641 & 0.0238 & -0.0697,0.1173 & -0.0234 & -0.1233,0.0766  \\ 
 $E[Y\{1,M(1)\}]-[Y\{0,M(1)\}]$ &  -0.0145 & -0.0657,0.0368 & -0.0013 & -0.1511,0.1486 & -0.0351 & -0.1653,0.0951 \\ 
 $E[Y\{1,M(1)\}]-[Y\{1,M(0)\}]$ &  -0.0978 & -0.1547,-0.0408 & -0.0270 & -0.0573,0.0033 & -0.0077 & -0.0289,0.0135 \\ 
$E[Y\{0,M(1)\}]-[Y\{0,M(0)\}]$ &  0.0113 & -0.0392,0.0618 & -0.0019 & -0.0935,0.0896 & 0.0040 & -0.0269,0.0349 \\ 
   \hline
\end{tabular}
\end{adjustbox}
\label{apptab}
\end{table}

%age, gender, ethnicity, language competency, education, marital
%status, household size and income, previous receipt of social aid, and family background (e.g., parents’ education),
%as well as health and health-related behavior at baseline.

%Describe the study background and theoretical rationale for investigating the mechanisms of interest. Include supporting evidence or theoretical rationale for why the intervention or exposure might have a causal relationship with the proposed mediators. Include supporting evidence or theoretical rationale for why the mediators might have a causal relationship with the outcomes

%We apply ourmethod to the Job Corps study, which was conducted in themid-1990s to assess the publicly funded US JobCorps program and used an experimental design in which access to Job Corps was assigned at random. The Job Corps

\end{document}